\documentclass[fleqn,usenatbib]{mnras}

\usepackage{newtxtext,newtxmath}

\usepackage[T1]{fontenc}
\usepackage{ae,aecompl}

\usepackage[utf8]{inputenc}

\usepackage{graphicx}	
\usepackage{amsmath}	
\usepackage{amssymb}	
\usepackage{bm}
\usepackage{color}
\usepackage{soul}

\newcommand\fq{K}
\newcommand\fpsi{ \Psi}
\newcommand\G{ \mathcal{G}}

\title[Asteroseismology of core-collapse supernovae - II]{Towards asteroseismology of core-collapse supernovae with gravitational-wave observations - II. Spacetime perturbations}

\author[A.~Torres-Forn\'e, P.~Cerd\'a-Dur\'an, A.~Passamonti, M.~Obergaulinger and Jos\'e A.~Font]
{Alejandro Torres-Forn\'e$^{1}$\thanks{E-mail: alejandro.torres@uv.es},
Pablo Cerd\'a-Dur\'an$^{1}$,
Andrea Passamonti$^{2}$, 
\newauthor
Martin Obergaulinger$^{1}$   and Jos\'e A. Font$^{1,3}$
\\
$^{1}$Departamento de Astronom\'ia y Astrof\'isica, Universitat de Val\`encia, C/ Dr. Moliner, 50, Burjassot (Valencia) E46100, Spain. \\
$^{2}$Departamento de F\'isica Aplicada, Universitat d'Alacant, Ap. Correos 99, E03080 Alicante, Spain. \\
$^{3}$Observatori Astron\`omic, Universitat de Val\`encia, C/ Catedr\'atico Jos\'e Beltr\'an 2, 46980, Paterna (Val\`encia), Spain.
}

\date{Accepted XXX. Received YYY; in original form ZZZ}

\pubyear{2017}

\begin{document}
\label{firstpage}
\pagerange{\pageref{firstpage}--\pageref{lastpage}}
\maketitle

\begin{abstract}
Improvements in ground-based, advanced gravitational wave (GW) detectors may allow in the near future to observe the GW signal
of a nearby core-collapse supernova. For the most common type of progenitors, likely with slowly rotating cores, the dominant GW emission mechanisms 
are the post-bounce oscillations of the proto-neutron star (PNS) before the explosion. We present a new procedure to compute the eigenmodes of the system formed by the PNS and the stalled accretion shock in general relativity including spacetime perturbations. The new method improves on previous results by accounting for perturbations of both the lapse function and the conformal factor. We apply our analysis to two numerical core-collapse simulations and show that our  
improved method is able to obtain eigenfrequencies that accurately match the features observed in the GW signal and to predict the qualitative behaviour 
of quasi-radial oscillations. Our analysis is possible thanks to a newly developed algorithm to classify the computed 
eigenmodes in different classes (f-, p-, and g-modes), improving previous results which suffered from misclassification issues. 
We find that most of the GW energy is stored in the lowest order eigenmodes, in particular in the $^2g_1$ mode and in the $^2f$ mode.
Our results also suggest that a low-frequency component of the GW signal attributed in previous works to the characteristic frequency of the Standing Accretion Shock Instability, should be identified as the fundamental quadrupolar f-mode. We also develop a formalism to estimate the contribution of quasi-radial ($l=0$) modes to the quadrupolar component of the GW emission in the case of a deformed background, with application to rapidly rotating cores. 
This work provides further support for asteroseismology of core-collapse supernovae and the inference of PNS properties based on GW observations.
\end{abstract}



\newcommand{\runi}{{\hat {\boldsymbol r}}}
\newcommand{\thetauni}{{\hat {\boldsymbol \theta}}}
\newcommand{\varphiuni}{{\hat {\boldsymbol \varphi}}}

\section{Introduction}

With the LIGO/Virgo discovery of gravitational waves (GWs) from a binary neutron star (BNS) merger and the subsequent follow-up observations across the electromagnetic spectrum by dozens of astronomical facilities and several neutrino telescopes, the field of multi-messenger astronomy has started~\citep{MMA}.  Significant advances toward explaining a number of open issues in relativistic astrophysics have been made thanks to the single observation of GW170817, from the mechanism behind short gamma-ray bursts~\citep{Monitor:2017mdv} to the $r$-process-mediated
nucleosynthesis of heavy elements in kilonovae~\citep{Abbott:2017wuw}, along with independent measures of cosmological parameters~\citep{Abbott:2017xzu}. Moreover, GWs from BNS mergers offer the opportunity to probe the properties high-density matter, yielding constraints on the neutron star radii and on the equation of state (EOS)~\citep{Annala,Fattoyev,De,Abbott:radii}.

A type of transient GW signal that remains to be detected is the one produced in core-collapse supernovae (CCSNe), associated with the catastrophic collapse of the unstable iron cores that massive stars (from $\sim 8 M_\odot$ to $\sim 100  M_\odot$) develop at the end of their cycles of thermonuclear burning reactions. Collapsing stars produce rich and complex gravitational waveforms. When detected, those may provide  important information about the phenomenology of the scenario, specially when combined with observations of their electromagnetic emission and neutrino emission. Their typical range of frequencies falls within the most sensitive region of the advanced LIGO and Virgo detectors. CCSNe are rare events, happening at rates of about one per century~\citep[see e.g.][]{Gossan:2016} observable within the Milky Way. This may explain why GWs from CCSNe have not yet been detected.  In terms of signal amplitudes, it would be possible to detect magneto-rotational explosions up to $\sim 5$ Mpc \citep{Gossan:2016}, but the event rate for such class of explosions is even lower,  $\sim 10^{-4} \rm{yr}^{-1}$
\citep[less than $1\%$ of all CCSNe, see discussion in ][]{Woosley:2006b}. Despite this constraint, there have been attempts to detect the GW emission of supernova signals with the current network of detectors, either through techniques that search generic short-duration burst signals  without targeting specific supernova events or  with targeted  searches that use the  known sky-location and  time window of electromagnetic or neutrino events~\citep{Thrane:2015,Klimenko:2016}.

Unlike the case of BBH mergers, it is currently not possible to relate uniquely and unambiguously the properties of the progenitor stars (such as mass, rotation rate, metallicity, or magnetic fields) with the resulting waveforms. The reasons are diverse: (i) the complex non-linear dynamics associated with the evolution of a fluid interacting with neutrino radiation, (ii) the stochastic and chaotic behaviour of instabilities (both during and prior to the collapse of the star), (iii) the uncertainties in the stellar evolution of massive stars (specially regarding the treatment of convection, magnetic fields and angular momentum transport), and (iv) the uncertainties in the nuclear and weak interactions necessary for the high-density EOS and neutrino radiation, respectively.

In CCSN, GWs are produced during the hydrodynamical bounce and during the evolution of the instabilities that occur in the cavity formed by the newborn neutron star and the accretion shock. The violent dynamics excite different modes of oscillation of the proto-neutron star (PNS) and of its surroundings, including the shock wave. 
Most of the previous work on the field
\citep{Reisenegger:1992,Ferrari:2003,Ferrari:2004,Passamonti:2005,Kruger:2015,Sotani:2016,Camelio:2017}
has focused in the computation of the oscillations modes of the PNS alone (without the interaction
with the shock) and with a simplified background PNS model.
In a previous paper~\citep{Torres-Forne:2018} we studied the relationship between the spectrum of oscillations of the PNS - shock wave system after core bounce (analyzing linear perturbations in general relativity of this background system) and the characteristic GW frequencies  obtained in numerical simulations.  In particular, we showed that both spectra are closely related, which can be used to identify the oscillation modes of the PNS, such as the $g$-modes and the $p$-modes, obtaining a remarkable correspondence with the time-frequency distribution of the GW signal. This result showed that it is possible to perform CCSN asteroseismology and serves as a starting point to carry out inference of astrophysical parameters of PNS using the information contained in the gravitational waveforms to be detected in future observations of current or third-generation GW detectors.
We note that these results have been recently extended by \cite{Morozova:2018}, who included spacetime perturbations in their analysis, albeit only of the lapse function, showing that spacetime perturbations have indeed an impact in the calculation.

In the current paper, we build on the approach presented in~\citet{Torres-Forne:2018} and incorporate to the methodology an augmented set of spacetime perturbation equations with respect to~\citet{Morozova:2018}. The aim of this work is to understand the features observed in the GW signal of CCSN simulations, which have been interpreted as $g$-modes of the PNS \citep{Murphy:2009,Mueller:2013,Cerda-Duran:2013,Yakunin:2015,Kuroda:2016, Andresen:2017}. It also attempts to shed some light on the imprint of the standing accreting shock instability \citep[SASI,][]{Blondin:2003,Foglizzo:2007} in the GW signal observed in numerical simulations~\citep{Cerda-Duran:2013,Kuroda:2016,Andresen:2017}. The paper is organised as follows: In Section~\ref{sec:linear_pert} we develop our linear analysis formalism and estimate the GW emission for the case of a background deformed by rotation. In Section~\ref{sec:eigen} we present the numerical methods used to solve the eigenvalue problem and the algorithms used in the classification of the eigenmodes. Section~\ref{sec:simulations}  presents our results and the comparison with the GW signal computed in the numerical simulations. Finally, in Section~\ref{sec:summary} we summarise our results and discuss their implications. The paper contains three appendices with technical details regarding numerical aspects. Throughout this paper we use a spacelike metric signature $(-,+,+,+)$ and $c = G = 1$ (geometrised) units, where $c$ stands for the speed of light and $G$ is Newton’s gravitational constant.  As customary, Greek indices run from 0 to 3, Latin indices from 1 to 3, and we use Einstein’s summation convention for repeated indices.

\section{Linear perturbations of a spherically-symmetric background}
\label{sec:linear_pert}

We start our analysis with the description of the perturbations of a spherically-symmetric, self-gravitating, equilibrium configuration. 
 The interested reader is addressed to \citet{kokkotas:1999} and to \citet{friedman:2013} for detailed
information on linear perturbations of compact stars and asteroseismology. Classically, this analysis was performed in Schwarzshild coordinates by~\citet{Thorne:1967}. In our work we use isotropic coordinates instead, which are closer to the gauge condition used in the relativistic CCSN numerical simulation we employ for the comparison~\citep{Cerda-Duran:2013}, which is based on the conformally-flat approximation~\citep{Isenberg08,Wilson96}. Moreover, the derivation of the equations in these coordinates also bears resemblance with the equations in the Newtonian case \citep[see][]{Reisenegger:1992}, which makes it easier to identify the role of the different terms in the equations and to interpret the solutions. This choice of gauge also makes it straightforward to perform the mode analysis of the Newtonian simulations we make use of~\citep{OJA2018}.

The metric in a $3+1$ foliation of the space-time in coordinates $(t,x^i)$ is
\begin{equation}
ds^2 = g_{\mu\nu}dx^\mu dx^\nu = (\beta^i\beta_i-\alpha^2) dt^2 + 2 \beta_i dt dx^i + \gamma_{ij} dx^i dx^j,
\end{equation}
where $\beta^i$ is the shift 3-vector, $\alpha$ is the lapse function, and $\gamma_{ij}$ is the spatial 3-metric. 

In this work we consider the conformally flat condition approximation \citep[CFC ][]{Isenberg08,Wilson96} of general relativity (GR).
 This approximation has some advantages for simulations of CCSNe: 
i) In spherical symmetry the CFC metric is exact and it is equivalent to choosing maximal slicing and isotropic coordinates;
ii) In the post-Newtonian regime ($M/R<1$) it deviates from GR as first post-Newtonian correction times the ellipticity squared \citep{Kley:1999,Cordero-Carrion:2009};
iii) Direct comparisons with full GR simulations of the collapse of rapidly rotating stellar cores have shown an excellent agreement  in both
the dynamics and the GW signal~\citep{Shibata:2004,Ott:2007a,Ott:2007b}. Disagreement with GR has been estimated to be below $1\%$
in those cases \citep{Cerda-Duran:2005};
iv) The resulting equations can be applied directly to the analysis of numerical simulations in the CFC approximation and, with minor modifications, to Newtonian simulations. 
v) In the Cowling approximation the coordinates coincide with those used in \cite{Torres-Forne:2018}, making possible a direct comparison with our previous results.

In CFC, the spatial 3-metric is conformally flat,  $\gamma_{ij} = \psi f_{ij}$, 
where $\psi$ is the conformal factor and $f_{ij}$ is the spatial 3-metric. The Einstein equations
for the CFC metric form a purely elliptic system given by
\begin{align}
\nabla^2 \psi &= -2\pi \psi^5 \left ( E + \frac{K_{ij} K^{ij}}{16 \pi} \right ), \label{eq:CFC1}\\
\nabla^2 Q &= 2\pi Q \psi^4 \left ( E + 2 S + \frac{7 K_{ij} K^{ij}}{16 \pi} \right ), \label{eq:CFC2} \\
\nabla^2 \beta^i + \frac{1}{3} \nabla^i \nabla_j \beta^j&= 16 \pi \alpha \psi^4 S^i + 2 \psi^{10} K^{ij} \nabla_j \left (\frac{\alpha}{\psi^{6}}\right ) \label{eq:CFC3},
\end{align}
where  $\nabla^2$ and $\nabla_i$ are the Laplacian and nabla operators with respect to the flat 3-metric, respectively,
$K^{ij}$ is the extrinsic curvature, and $Q\equiv \alpha\psi$. The energy-momentum content couples
to the spacetime geometry through the projections of the energy-momentum tensor, $T_{\mu\nu}$, onto the $3+1$ foliation
\begin{align}
E &\equiv \alpha^2 T^{00}, &\quad
S_i &\equiv - \alpha^{-1}(T_{0i} - T_{ij} \beta^j), \\
S_{ij} & \equiv T_{ij}, &\quad
S & \equiv S_{ij} \gamma^{ij}.
\end{align}
We consider a perfect fluid, for which the energy-momentum tensor is given by
\begin{equation}
T^{\mu\nu} = \rho h u^{\mu} u^{\nu} + P g^{\mu\nu} \,,
\end{equation}
where $\rho$ is the rest-mass density, $P$ is the pressure, $u^{\mu}$ is the 4-velocity,
$h\equiv 1 + \epsilon + P/\rho$ is the specific enthalpy, and $\epsilon$ is the specific internal
energy. It is useful to define the energy density as $e \equiv \rho (1 + \epsilon)$.

The mass-conservation (continuity) and momentum conservation equations in GR read~\citep{Banyuls:1996}:
\begin{align}
\frac{1}{\sqrt{\gamma}} \partial_t \left [ \sqrt{\gamma}D \right ] + \frac{1}{\sqrt{\gamma}} \partial_i \left [ \sqrt{\gamma} D v^{*i} \right]
 &= 0, \label{eq:continuity} \\
\frac{1}{\sqrt{\gamma}} \partial_t \left [ \sqrt{\gamma} S_j \right ] 
+ \frac{1}{\sqrt{\gamma}} \partial_i \left [ \sqrt{\gamma} S_j v^{*i} \right ]
+ \alpha \partial_j P 
&= \frac{\alpha \rho h}{2} u^{\mu}u^{\nu} \partial_j g_{\mu\nu}, \label{eq:momentum} 
\end{align}
where $\gamma$ is the determinant of the three-metric. The conserved quantities are $D = \rho W$ and $S_j = \rho h W^2 v_j$, 
%
where $W=1/\sqrt{1-v_iv^i}$ is the Lorentz factor. 
The Eulerian and ``advective'' velocities are, respectively, 
\begin{align}
v^i & = \frac{u^i}{W} +\frac{\beta^i}{\alpha},\quad & \quad v^{*i} &= \frac{u^i}{u^0} =\alpha v^i - \beta^i.
\end{align}

Let us consider a solution of the hydrodynamics equations that is in equilibrium ($\partial_t=0$) and is static ($v^i=0$). 
In this case the metric equations (Eqs.~\ref{eq:CFC1}-\ref{eq:CFC3}) and the hydrodynamics equations (Eqs.~(\ref{eq:continuity})-(\ref{eq:momentum})) 
reduce to
\begin{align}
\nabla^2 \psi = -2\pi\psi^5 E, \label{eq:EQ1}\\
\nabla^2 Q = 2\pi Q \psi^4 (E+2S) \label{eq:EQ2}, \\
\frac{1}{\rho h} \partial_i P = -\partial_i \ln \alpha \equiv \G_i, \label{eq:G}
\end{align}
and $\beta^i=0$.
In the Newtonian limit $\G_i$ is the gravitational acceleration, whose only non-zero component is $\G_r\equiv \G$. 
The solution of Eqs.~(\ref{eq:EQ1})-(\ref{eq:G}) corresponds to the unperturbed state or background solution.

Following \cite{Torres-Forne:2018}, we consider linear adiabatic perturbations of the hydrodynamics equations with respect to the background equilibrium 
configuration. We use the same notation and denote  the Eulerian perturbations of the different quantities with $\delta$ (e.g. $\delta \rho$) and without it for the background quantities (e.g. $\rho$). Our previous work was based on the Cowling approximation. Here, we consider the general case including also the
metric perturbations, namely $\delta\alpha$, $\delta\psi$, and $\delta \beta^j$. For a weak gravitational field ($M/R\ll 1$), 
the leading term in the post-Newtonian expansion of the metric contributes only to $\alpha$, while the first correction to $\psi$ and $\beta^i$ 
appears at the first post-Newtonian level~\citep{Blanchet:1990}. Therefore, for the mildly relativistic gravitational field of a PNS, we expect $\delta \alpha$
to be dominant in front of $\delta \psi$ and $\delta \beta^i$. In this work we consider $\delta \beta^i=0$, which simplifies the equations
significantly. Although this approach does not fully include the metric perturbations, it allows us to assess the effect of 1PN corrections 
with respect to a ``Newtonian-like" approach in which only $\delta \alpha$ is considered, as in~\citet{Morozova:2018}.

We denote as $\xi^i$ the Lagrangian displacement of a fluid element with respect to its position at rest. Its value is related
to the advective velocity as
\begin{equation}
\partial_t \xi^i = \delta v^{*i}.
\end{equation}
The Lagrangian perturbation of any quantity, e.g. $\rho$, is related to the Eulerian perturbations as
\begin{equation}
\Delta \rho = \delta \rho + \xi^i\partial_i \rho.
\end{equation}
The linearised version of Eqs.~(\ref{eq:continuity}) and (\ref{eq:momentum}) are
\begin{align}
\frac{ \Delta \rho }{\rho} &= - \left( \partial_i \xi^i + \xi^i \partial_i \ln \sqrt{\gamma} \right) - 6 \frac{\delta \psi}{\psi} \, , \label{eq:cnt} \\
\rho h \, \partial_t \delta v_i + \alpha \partial_i \delta P &= - \delta \left( \rho h \right) \partial_i \alpha - \rho h (\partial_i \delta \alpha - \delta \alpha \partial_i \ln \alpha) \,. \label{eq:meq2}
\end{align}
We use spherical coordinates, $\{r,\theta,\varphi\}$, in which $\sqrt{\gamma} = \psi^6 r^2 \sin\theta$.
The condition of adiabaticity of the perturbations implies that 
\begin{equation}
\frac{\Delta P}{\Delta \rho} = \left . \frac{\partial P}{ \partial \rho} \right |_{\rm adiabatic} = h c^2_{{\rm s}} = \frac{P}{\rho}\Gamma_1, \label{eq:adiab}
\end{equation}
where $c_{\rm s}$ is the relativistic speed of sound and $\Gamma_1$ is the adiabatic index.
This allows us to write
\begin{align}
\delta \left( \rho h \right) = \left( 1 + \frac{1}{c_s^2} \right) \delta P - \rho h \, \xi^i \mathcal{B}_{i} \, , \label{eq:rh}
\end{align} 
where
\begin{equation}
\mathcal{B}_i \equiv \frac{\partial_i e}{\rho h} - \frac{1}{\Gamma_1}\frac{\partial_i P}{P},
\end{equation}
is the relativistic version of the Schwarzschild discriminant. Since the background is spherically symmetric, 
the only non-zero component is ${\mathcal B}_r\equiv {\mathcal B}$.

The radial and angular parts of Eq.~(\ref{eq:meq2}) are given by 
\begin{align}
& \rho h \, \psi^4 \alpha^{-2} \frac{ \partial^2 \xi^r }{\partial t^2 } + \partial_r \delta P = \delta \left( \rho h \right) \G 
- \rho h \alpha^{-1} (\partial_r \delta \alpha + \delta \alpha \G),  \label{eq:vr} \\
& \rho h \, \psi^4 \alpha^{-2} r^2 \frac{ \partial^2 \xi^{\theta} }{\partial t^2 } + \partial_{\theta} \delta P = 
- \rho h \alpha^{-1} \partial_\theta \delta \alpha, \label{eq:vth} \\
& \rho h \, \psi^4 \alpha^{-2} r^2 \sin^2\theta\frac{ \partial^2 \xi^{\varphi} }{\partial t^2 } + \partial_{\varphi} \delta P = 
- \rho h \alpha^{-1} \partial_\varphi \delta \alpha, \label{eq:vph} 
\end{align} 
where we have used that, in the coordinate basis, the covariant components of the velocity are given by $\delta v_r = \psi^4 \delta v^r$, $\delta v_{\theta} = r^2 \psi^4 \delta v^{\theta}$
and $\delta v_{\varphi} = r^2 \sin^2 {\theta} \psi^4 \delta v^{\varphi}$.

Additionally we need equations for the metric perturbations $\delta \hat Q$ and $\delta \hat \psi$, 
that can be obtained by linearising Eqs.~(\ref{eq:CFC1})-(\ref{eq:CFC2}) and subtracting the 
background solution (Eqs.~(\ref{eq:EQ1})-(\ref{eq:EQ2})),
\begin{align}
\nabla^2 \delta\psi =& -10\pi e \psi^4 \delta\psi - 2 \pi \psi^5 \left [ \frac{\delta P}{c_s^2} - \rho h \xi^r \mathcal{B} \right ],  \label{eq:mpert1} \\
\nabla^2 \delta Q =& 2 \pi (\rho h + 5 P) \psi^4 (\delta Q + 4 \alpha \delta \psi) \nonumber \\
&+ 2\pi \alpha \psi^5 \left [  \left ( 6 + \frac{2}{c_s^2} \right ) \delta P -  \rho h \xi^r \mathcal{B} \right ], \label{eq:mpert2} 
\end{align}
where we have used that $\delta K^{ij} = 0$, for $\delta \beta^i = 0$.

We perform an expansion of the perturbations with a harmonic time dependence of frequency $\sigma$ and a spherical-harmonic 
expansion for the angular dependence
\begin{align}
\delta P = \delta \hat{P} \,\,& Y_{lm} e^{- i \sigma t}, \\
{\bf \xi}^r = \eta_r \,\, &Y_{lm} e^{- i \sigma t} , \\
{\bf \xi}^\theta = \eta_\perp \,\,& \frac{1}{r^{2}} \partial_\theta Y_{lm} e^{- i \sigma t}, \\
{\bf \xi}^\varphi = \eta_\perp \,\,& \frac{1}{r^2 \sin^2\theta} \partial_\varphi Y_{lm} e^{- i \sigma t},
\end{align}
and all scalar quantities (e.g. $\delta Q$ and $\delta \psi$) in a equivalent way to $\delta P$.
The quantities $\eta_r, \eta_{\perp}$ and the scalar perturbations with the hat (e.g. $\delta \hat P$, $\delta \hat Q$ and $\delta \hat \psi$), depend on the 
radial coordinate only. 

\subsection{Non-radial perturbations ($l\ne 0$)}
\label{sec:l2}

For $l\neq0$, by inserting the spherical-harmonic expansion into equations (\ref{eq:vr})-(\ref{eq:vth}) we obtain:
\begin{align}
& - \sigma^2 q\, \eta_r + \partial_r \delta \hat P = \delta \widehat{\rho h}  \G 
- \rho h \alpha^{-1} (\partial_r \delta\hat\alpha + \G \delta\hat\alpha), \label{eq:vr2} \\
& - \sigma^2 q \,\eta_{\perp} + \delta \hat P = 
-\rho h \alpha^{-1}\delta\hat\alpha, \label{eq:vth2} 
\end{align} 
where for convenience we have defined $q \equiv \rho h \, \alpha^{-2} \psi^4$.
From Eq.~(\ref{eq:vth2}) it follows that
\begin{equation}
\delta \hat P = q \sigma^2 \eta_\perp \, -\frac{\rho h}{\alpha} \delta\hat\alpha = 
q \sigma^2 \eta_\perp \, +\rho h \left (  \frac{\delta\hat\psi}{\psi} -\frac{\delta \hat Q}{Q} \right ). \label{eq:dp}
\end{equation}
Using Eqs.~(\ref{eq:dp}) and (\ref{eq:rh}) to simplify Eqs.~(\ref{eq:cnt}) and (\ref{eq:vr2}) we obtain
\begin{align}
\partial_r \eta_{r} 
+ \left[ \frac{2}{r} + \frac{1}{\Gamma_1}\frac{ \partial_r P }{P } + 6 \frac{ \partial_r \psi }{\psi}  \right] \eta_r 
+ \frac{\psi^4 }{\alpha^2 c^2_s} \left( \sigma^2 - \mathcal{L}^2 \right) \eta_\perp  \nonumber \\
= \frac{1}{c_s^2} \frac{\delta \hat Q}{Q} - \left ( 6 + \frac{1}{c_s^2} \right )\frac{\delta\hat\psi}{\psi}
 \, , \label{eq:er} \\
\partial_r \eta_\perp 
- \left( 1 - \frac{ \mathcal{N}^2 }{\sigma^2} \right) \eta_r 
+ \left[ \partial_r \ln q - \G \left( 1 + \frac{1}{c^2_s} \right) \right] \eta_\perp  \nonumber \\
= \frac{\alpha^2}{\psi^4\sigma^2} \left [ \partial_r (\ln \rho h) - \left ( 1 + \frac{1}{c_s^2} \right )\G \right ] \left ( \frac{\delta \hat Q}{Q} - \frac{\delta \hat \psi}{\psi}\right )
\, , \label{eq:eth} 
\end{align}
where $\mathcal{L}$ is the relativistic Lamb frequency defined as
\begin{equation}
\mathcal{L}^2 \equiv \frac{ \alpha^2 }{\psi^4} c^2_s \frac{l\left(l+1\right)}{r^2} \, ,
\end{equation}
and $ \mathcal{N}$ is the relativistic Brunt-V\"ais\"al\"a frequency 
\begin{equation}
 \mathcal{N}^2 \equiv \frac{ \alpha^2 }{\psi^4 } \G^i \mathcal{B}_i = \frac{\alpha^2}{\psi^{4}} \mathcal{B} \G .
\end{equation}

Eqs.~(\ref{eq:mpert1})-(\ref{eq:mpert2}) for the metric perturbations can also be simplified using Eq.~(\ref{eq:dp}).
After the spherical harmonic expansion they read
\begin{align}
\hat \nabla^2 \delta \hat \psi &= -2\pi \psi^5 \left [ \left ( 5 e + \frac{\rho h}{c_s^2} \right ) \frac{\delta\hat\psi}{\psi}
- \frac{\rho h}{ c_s^2} \frac{\delta \hat Q}{Q} \right ] \nonumber  \\
&- 2 \pi \rho h \psi^5 \left ( \frac{\psi^4 \sigma^2}{\alpha^2 c_s^2}\eta_\perp  - \mathcal{B} \eta_r \right ), \label{eq:metric:L1} \\
\hat \nabla^2 \delta \hat Q &= 2 \pi (\rho h + 5 P) \alpha \psi^5 (\frac{\delta \hat Q}{Q} + 4 \frac{\delta \hat \psi}{\psi}) \nonumber \\
& + 2\pi \rho  h \alpha \psi^5 \left [  \left ( 6 + \frac{1}{c_s^2} \right ) \left ( \frac{\psi^4\sigma^2}{\alpha^2 } \eta_\perp - \frac{\delta \hat Q}{Q} + \frac{\delta \hat \psi}{\psi} \right )
-  \eta_r \mathcal{B} \right ], \label{eq:metric:L2}
\end{align}
where 
\begin{equation}
\hat \nabla^2 \equiv \partial_{rr} + \frac{2}{r}\partial_r - \frac{l(l+1)}{r^2}.
\end{equation}
The system of Eqs.~(\ref{eq:er}), (\ref{eq:eth}), (\ref{eq:metric:L1}) and (\ref{eq:metric:L2}) can be reduced to a system of $6$ first-order ODEs
by introducing the variables $\fq \equiv \partial_r\delta\hat Q$ and $\fpsi \equiv \partial_r \delta\hat\psi$. The resulting system of equations can be written
in compact form as
\begin{equation}
\partial_r \bm u = \bm A \bm u, \label{eq:compact}
\end{equation}
where $\bm u \equiv (\eta_r,\eta_\perp, \fq,\delta\hat Q,\fpsi,\delta\hat\psi)^{T}$ and $\bm A$ is a $6\times 6$ coefficient matrix whose components are explicitly
written in appendix~\ref{sec:appendix:coeff}.

\subsection{Radial perturbations ($l=0$)}
\label{sec:l0}

In non-rotating stars, the Lagrangian displacement of radial oscillations has $\xi^\theta=\xi^\varphi=0$, i.e. $\eta_\perp=0$. This case cannot be 
treated as a particular case of the general derivation for $l\ne 0$ (previous section).  In this case, the continuity and momentum equations are
\begin{align}
\frac{ \delta \hat {\rho} }{\rho} = -\partial_r \eta_r - \left( \frac{2}{r}  +\frac{\partial_r \rho}{\rho} + 6 \frac{\partial_r \psi}{\psi} \right) \eta_r - 6 \frac{\delta \hat {\psi}}{\psi}, \label{eq:sph:rho}\\
-\sigma^2 q \eta_r + \partial_r \delta \hat{P} =   \G \delta\widehat{\rho h}   
- \rho h \alpha^{-1} (\partial_r \delta \hat{\alpha} + \delta \hat{\alpha} \,  \G) .  \label{eq:sph:v} 
\end{align}
From Eqs.~(\ref{eq:adiab})  and (\ref{eq:rh}) the various scalar perturbations are given by 
\begin{align}
\delta \hat{P} = P \Gamma_1 \left [ \frac{\delta \hat{\rho}}{\rho} +\eta_r \left ( \frac{\partial_r \rho}{\rho} - \frac{\partial_r P}{\Gamma_1 P}\right )\right ], \\
\delta \widehat{\rho h} = \left( 1 + \frac{1}{c_s^2} \right)  \delta \hat{P} - \rho h \,\eta_r \, \mathcal{B} . \label{eq:sph:rhoh}
\end{align}
Combining Eqs.~(\ref{eq:sph:rho})-(\ref{eq:sph:rhoh}) we obtain the following equations for $\eta_r$ and $\delta\hat P$
\begin{align}
\partial_r \eta_r &+ \left [ \frac{2}{r} +\frac{1}{\Gamma_1} \frac{\partial_r P}{ P}  + 6 \frac{\partial_r \psi}{\psi} \right ] \eta_r
+ \frac{1}{P\Gamma_1} \delta\hat P = - 6 \frac{\delta \hat \psi}{\psi}, \label{eq:L01}\\
\partial_r \delta\hat P &+ q (\mathcal{N}^2 - \sigma^2) \eta_r - \G \left (1 + \frac{1}{c_s^2} \right ) \delta\hat P
= -\rho h \alpha^{-1} (\partial_r \delta \hat \alpha + \delta \hat \alpha \G). \label{eq:L02}
\end{align}
Using Eqs.~(\ref{eq:mpert1}) and (\ref{eq:mpert2}) the metric perturbations are
\begin{align}
\hat \nabla^2 \delta\hat \psi =& -10\pi e \psi^4 \delta\hat \psi - 2 \pi \psi^5 \left [ \frac{\delta \hat P}{c_s^2} - \rho h \eta_r \mathcal{B} \right ],  \label{eq:metric:L01} \\
\hat \nabla^2 \delta \hat Q =& 2 \pi (\rho h + 5 P) \psi^4 (\delta \hat Q + 4 \alpha \delta \hat \psi) \nonumber \\
&+ 2\pi \alpha \psi^5 \left [  \left ( 6 + \frac{2}{c_s^2} \right ) \delta \hat P -  \rho h \eta_r \mathcal{B} \right ]. \label{eq:metric:L02} 
\end{align}
Similarly to the case with $l\ne0$, the system of Eqs.~(\ref{eq:L01})-(\ref{eq:metric:L02}) can be written as a system of $6$ first-order ODEs 
by using $\fq$ and $\fpsi$ and cast in the same compact form as Eq.~(\ref{eq:compact}), where in this case
$\bm u \equiv (\eta_r,\delta \hat P, \fq,\delta\hat Q,\fpsi,\delta\hat\psi)^{T}$ and $\bm A$ is a $6\times 6$ coefficient matrix whose components are explicitly
given in appendix~\ref{sec:appendix:coeff}.

\subsection{Boundary conditions}
\label{sec:bcs}

As in \cite{Torres-Forne:2018}, we impose zero-displacement boundary conditions at the shock location
\begin{equation}
\xi_r|_\textrm{shock} = 0 \quad \to \quad \eta_r|_\textrm{shock} = 0,
\end{equation}
which is a consequence of the impossibility of perturbations to propagate across the shock from the subsonic
to the supersonic region.
At the origin ($r=0$) we impose regularity, which for $l\ne0$ \citep[see][]{Reisenegger:1992}
results in\footnote{Note that in \cite{Torres-Forne:2018} there is a missing $r$ factor in the text (but not in the calculations).}
\begin{equation}
\eta_r|_{r=0} = \frac{l }{r} \eta_\perp|_{r=0} \propto r^{l-1}, \label{eq:regularity}
\end{equation}
and for $l=0$ in
\begin{align}
& \eta_r|_{r=0} \propto r \label{eq:regularity0}, \\
& \delta\hat P|_{r=0} =  - P \Gamma_1 \left [ 6 \frac{\delta \hat \psi}{\psi}  + \frac{\eta_r}{r} \right ]_{r=0}. \label{eq:BC:L0}
\end{align}
The latter expression has been obtained by evaluating Eq.~(\ref{eq:L01})
at $r=0$ and imposing the boundary condition for $\eta_r$ given by Eq.~(\ref{eq:regularity0}). 

Regarding metric perturbations, regularity at $r=0$ implies that 
\begin{align}
\fq|_{r=0} = \frac{l}{r}   \delta \hat Q|_{r=0}&\propto r^{l-1} \quad &; \quad \fpsi|_{r=0} = \frac{l}{r} \delta \hat \psi|_{r=0} &\propto r^{l-1}.
\end{align}
As a consequence all four metric functions are zero at $r=0$ for $l\ge 2$.

Outside the sources ($\rho = 0$) the metric perturbations fulfil  $\hat \nabla^2 \delta\hat \psi =0$ and  $\hat \nabla^2 \delta\hat Q=0$
and hence the solution decays as $r^{-(l+1)}$. Outside the shock, density is non zero but, since $\eta_r = \eta_\perp=0$,  
and $\delta \hat Q$ and $\delta\hat \psi$ decay radially, it can be shown that, sufficiently far away from the shock (but still inside 
the star), the solution also decays as $r^{-(l+1)}$. For simplicity and given that outside the shock there is a considerable drop in
density, we impose this behaviour at the shock location. As a consequence
\begin{align}
[\fq + (l+1) \delta \hat Q / r ]_{\rm shock} = 0, \label{eq:BCS1} \\
[\fpsi  + (l+1) \delta \hat \psi / r ]_{\rm shock} = 0. \label{eq:BCS2}
\end{align}

We note that these boundary conditions differ with respect to those used by~\cite{Morozova:2018}, which impose boundary conditions 
at the PNS surface, assuming it is a free surface ($\Delta P = 0$).

\subsection{Gravitational wave emission}

\subsubsection{Spherical background}

Let us consider a general linear perturbation of the spherically-symmetric background
considered in the previous sections as a combination of eigenfunctions, which we denote hereafter as
$\delta \rho_{lm}$ (and so forth for other perturbed quantities):
\begin{equation}
\delta\rho = \sum_{l=2}^{\infty}\sum_{m=-l}^{+l} \delta \rho_{lm}
= \sum_{l=2}^{\infty} \sum_{m=-l}^{+l} \delta \hat \rho_{lm}  \,Y_{lm},
\end{equation}
where  $\delta \hat \rho_{lm}$ can be computed as
\begin{equation}
\delta{\hat\rho}_{lm} \approx \rho \left ( 
\frac{\mathcal{N}^2}{\G}\eta_r^{lm} + \frac{\sigma^2}{c_s^2}\eta^{lm}_\perp \right ).
\label{eq:rho1}
\end{equation}
In the Newtonian limit, the energy stored in all the $(l, m)$ modes with a certain amplitude can be approximated as
\begin{equation}
E_{lm}= \frac{\sigma^2}{2}\int_0^{r_{\rm shock}}\mathcal{E}_{lm}~r^2 dr,\label{eq:energy2}
\end{equation}
where $\mathcal{E}_{lm}$ is the eigenmode energy density defined as
\begin{equation}
\mathcal{E}^{lm}(r) = \rho~ \left[{\eta_r^{lm}(r)}^2+l(l+1)\frac{{\eta_\perp^{lm}(r)}^2}{r^2} \right] .
\label{eq:energyden2}
\end{equation}
The $(l,m)$-mass multipole moment ($l\ge2$) can be defined as
\begin{equation}
\delta D_{lm} \equiv \int dV r^l \delta\rho \,Y^*_{lm} = \int_0^{r_{\rm shock}} r^{l+2} \delta \hat \rho_{lm}\, dr.
\label{eq:massmult}
\end{equation}
As a consequence, only the $(l,m)$ mode contributes to $\delta D_{lm}$. As we address
in the next section, this is a direct consequence of considering a spherically symmetric background.
Following~\citet{thorne:1969}, it is possible to compute the radiated power
in GWs as,
\begin{equation}
P_{lm} = \frac{G}{8\pi c^{2l+1}} 
\frac{(l+1)(l+2)}{(l-1)l}\left[
\frac{ 4\pi \sigma^{l+1}}{(2l+1)!!} \right]^2 \left ( |\delta D_{lm}|^2 + \frac{1}{c^2}|\delta J_{lm}|^2 \right),
\label{eq:power}
\end{equation}
where $\delta J_{lm}$ are the current multipoles and we show the factors $c$ and $G$ explicitly for the sake of the discussion. 

The dominant GW emission channel is the mass quadrupole ($l=2$). The contribution by higher-order mass multipoles and current multipoles is suppressed by at least a factor $1/c^2$ and is not relevant for GW detection. The implication is that, if the background is spherically symmetric,  
only $l=2$ oscillation modes are relevant for GW emission, and the radiated power of the relevant modes can be computed
as
\begin{equation}
P_{2m} = \frac{G}{c^{5}} 
\frac{4\pi\sigma^6}{75} |\delta D_{2m}|^2 .
\label{eq:power}
\end{equation}
The amplitude of the GW emitted for $(l,m)=(2, 0)$ modes can  be computed as 
\begin{align}
h_{+} &= -\frac{1}{D}\sin^2\Theta \frac{8\pi}{5} \sigma^2 \delta D_{20}\,, \\
h_{\times} &= 0 \,,
\label{eq:h+}
\end{align}
where $D$ is the distance to the source and $\Theta$ is the observation angle with respect to the symmetry axis of the mode. 

In order to compare the power emitted by different modes with $l=2$, we define the GW emission efficiency as
\begin{equation}
\textrm{(GW efficiency)}= \frac{P}{E f},
\label{eq:effi}
\end{equation}
where $f$ is the frequency of the mode. This equation gives an idea of the fraction of the mode energy radiated in GWs per oscillation cycle.

\subsubsection{Deformed background}
\label{sec:deformed}

One of the limitations of our linear analysis is that is only applicable if the background is spherically symmetric. However, in the collapse
of rapidly rotating cores, deformations in the star induced by rotation are known to enhance considerably the GW emission, increasing the wave amplitude 
by a factor $10-1000$ \citep[see e.g.][]{Fryer:2011}. Another motivation to consider a deformed background is the possible presence of $l=0$ (quasi-radial) modes in the GW signal. \cite{Cerda-Duran:2013} noted that some of the features observed in the GW signal of their simulation could be explained by a $l=0$ mode. As we show next, the presence of a deformed background allows for modes with $l\ne 2$ to contribute to the dominant GW channel ($l=2$).

Let us consider an axisymmetric background of the form
\begin{equation}
\rho(r,\theta) = \rho_0 (r)+ a \tilde\rho_2 (r) Y_{20} (\theta),
\end{equation}
where $\rho_0$ is the spherically-symmetric contribution to the background, $\tilde \rho_2$, whose volume integral is normalised to unity, 
gives the radial dependence of the quadrupolar deformation, and $a$ is a parameter controlling the amount of deformation 
($a=0$ corresponds to the spherically symmetric case considered above). We consider only quadrupolar deformations
of the background because those produce the strongest coupling between  modes with $l\ne 2$ and the quadrupolar component of the GW signal.
This allows us to compute the leading contribution to the signal. For simplicity, we consider $a\ll 1$. In particular we assume that
the quadrupolar deformations are sufficiently small to be neglected when compared to the perturbations themselves, i.e.
\begin{equation}
a \tilde \rho_2 << \delta \rho_{2m} << \rho_0.
\end{equation}
This allows to compute the leading-order contribution to the GW signal of modes with $l \ne 2$, without the added complexity of the linear analysis
including rotation.

Under these conditions, the equations for the linear perturbations are identical to the equations for the spherical background. This means that we can use
the eigenvalues and eigenfunctions already computed to estimate the leading-order contribution to the GW signal. The only necessary step
is to compute the contribution of modes with $l \ne 2$ to the mass quadrupole $\delta D_{2m}$.

Since $a\ll 1$, we can consider that the new background can be described as a continuous deformation of the spherically symmetric background given by
$\rho_0$. This deformation can be described by a displacement vector $X^i$  defined such that, for surfaces of constant density, corresponds to a
Lagrangian perturbation
\begin{equation}
a \tilde\rho_2 Y_{20} + X^i \partial_i \rho_0 = 0.
\label{eq:lag:def}
\end{equation}
Since the deformations is proportional to $Y_{20}$,  $X^i$ is purely radial, i.e. $X^r = N^r(r) Y_{20}$ and $X^\theta=X^\varphi=0$.
Note that this only happens for quadrupolar deformations of the background and simplifies significantly the analysis.
Substituting in Eq.~(\ref{eq:lag:def}) one arrives to
\begin{equation}
N^r (r) = -\frac{a \tilde\rho_2}{\partial_r \rho_0}. \label{eq:def:N}
\end{equation}
Let us consider a new set of coordinates $(r', \theta', \varphi')$ defined by
\begin{equation}
r' \equiv r + X^r = r + N^r Y_{20} \quad ; \quad
\theta' \equiv \theta, \quad ; \quad
\varphi' \equiv \varphi.
\end{equation}
By construction, in the new coordinates $\rho(r,\theta)=\rho(r')$, i.e. they are adapted to the deformation.
Using Eq.~(\ref{eq:def:N}) we can write
\begin{equation}
dr' = \left ( 1 + \partial_r N^r Y_{20} \right ) dr \quad ; \quad d\theta' = d\theta \quad ; \quad d\varphi' = d\varphi,
\end{equation}
which allows us to write the line element as
\begin{equation}
dl^2 \approx \left ( 1 - 2 \partial_r N^r Y_{20}\right ) dr'^2
+ r'^2 \left ( 1 - \frac{2 N^r Y_{20}}{r'} \right ) d \Omega'^2, 
\end{equation}
where we have used that $X^r/r \ll 1$ and we have neglected higher order corrections.
Here $d\Omega' \equiv \sin\theta' d\theta' d\varphi'$. 
Finally, the volume element in the new coordinates reads
\begin{equation}
dV \approx \left [ 1 - 2 \left ( \partial_r N^r + \frac{N^r}{r'} \right ) Y_{20} \right] r'^2 dr' d\Omega' .
\end{equation}

Therefore, using the fact that the eigenfunctions and eigenvalues are unchanged at this level of approximation, 
we can compute the mass quadrupole by considering the change in the volume element in Eq.~(\ref{eq:massmult}). 
The resulting contribution to  the mass quadrupole of a mode with $(l',m')$ is
\begin{align}
\delta D_{2m,l'm'} &= \int dV r^2 \delta\hat\rho \,Y_{l'm'} \,Y^*_{lm} \nonumber \\
&\approx 
\int  \left [ 1 - 2 \left ( \partial_r N^r + \frac{N^r}{r'} \right ) Y_{20} \right] r'^2 dr' d\Omega'  \, \delta\hat\rho \,Y_{l'm'} \,Y^*_{lm}.
\label{eq:massquad}
\end{align}
For oscillation modes with $l'=2$
\begin{equation}
\delta D_{2m,2m'} = \delta_{mm'}  \int_0^{r_{\rm shock}} r'^4 \, \delta\hat\rho_{2m'} dr' + \mathcal{O}\left(a \right), 
\end{equation}
which is equivalent to the case with spherical background given by Eq.~(\ref{eq:massmult}). However, for $l'\ne 2$, 
which gives a negligible contribution to the GW signal for the spherical background, we now obtain a non-zero contribution
given by
\begin{eqnarray}
\delta D_{2m,l'm'} &=& \int r'^4 dr' d\Omega' \left ( 2 \partial_r N^r + 4 \frac{N^r}{r'} \right ) \delta \hat \rho_{l'm'} Y_{20} Y_{l'm'} Y^*_{2m} 
\nonumber \\ &+& \mathcal{O}\left(a^2 \right), 
\end{eqnarray}
The angular part of the integral can be easily computed \citep{Arfken:1995}, and the only non-zero contributions
are 
\begin{align}
\delta D_{20,00} &= \delta \hat D_{00}, \label{eq:d00}\\
\delta D_{20,40} & = \frac{6}{7}\delta \hat D_{40},   \label{eq:d40}\\
\delta D_{2\pm1,4\pm1} & = \frac{\sqrt{30}}{7}\delta \hat D_{4\mp1},   \label{eq:d41}\\
\delta D_{2\pm2,4\pm2} & = \frac{\sqrt{15}}{7}\delta \hat D_{4\mp 2},  \label{eq:d42}
\end{align}
where 
\begin{equation}
\delta\hat D_{l'm'} \equiv  \frac{1}{\sqrt{4\pi}}\int_0^{r_{\rm shock}} r'^4 \left ( 2 \partial_r N^r + 4 \frac{N^r}{r'} \right ) \delta \hat\rho_{l'm'} dr'. \label{eq:dlm0}
\end{equation}
Note that these integrals are independent of $m'$, except for the mode amplitude, which can be different for each $m'$.
We can extract some conclusions from these results: i) the only oscillation modes contributing to the quadrupolar GW emission 
are those with $l'=0,2,4$, and ii) modes with $m'$ only produce GW in the $(2,m')$ channel. 
We will focus next on the contribution by $\delta D_{20,00}$ and $\delta D_{20,40}$, since the remaining contributions are just proportional
to the latter.

In order to use Eq.~(\ref{eq:d00}-\ref{eq:dlm0}) to compute the GW power of $l=2,4$ modes in a deformed background, one needs
to compute $\tilde \rho_2$ from the original multidimensional simulation. To simplify this process and develop our intuition about
the meaning of the corrections due to the deformation we will make some assumptions. Let us consider the deformation at a fixed radius $r$. 
The polar and equatorial radius of the density isocontour corresponding to $\rho_0(r)$ are given by:
\begin{align}
r_{\rm e} & = r + b, \\
r_{\rm p} & = r - 2b, \\
b &\equiv - \sqrt{\frac{5}{16 \pi}}N^r. \label{eq:bdef}
\end{align}
The ellipticity of the system at this radius, considering an oblate form ($r_{\rm p} < r_{\rm e}$) is
\begin{equation}
e \equiv \sqrt{1-\frac{r^2_{\rm p}}{r^2_{\rm e}}}= \frac{\sqrt{3 b (2 r - b)}}{r+b}. \label{eq:edef}
\end{equation}

To simplify further, we consider constant ellipticity at different radii. From Eqs.~(\ref{eq:bdef}) and (\ref{eq:edef})
this implies that
\begin{equation}
N^r =  - \sqrt{\frac{16 \pi}{5}} \frac{ (3 - e^2) - 3 \sqrt{1-e^2}}{3+e^2} r,
\end{equation}
such that, for $e=0$ (spherical) it results $N^r = 0$. Using this expression, the integral needed to
compute the contribution to the mass quadrupole is
\begin{equation}
\delta\hat D_{l'm'} \equiv  -\sqrt{\frac{4}{5}} \frac{ (3 - e^2) - 3 \sqrt{1-e^2}}{3+e^2}   \int_0^{r_{\rm shock}} r'^4 \delta \hat\rho_{l'm'} dr'. \label{eq:dlm}
\end{equation}
Note that the integral is the same as for $l=2$ modes, but using the corresponding multipole. 

\section{Eigenmode calculation}
\label{sec:eigen}

\subsection{Background models}

To test the capabilities of linear perturbation analysis we compute the eigenmodes for two CCSN 
simulations, for which we have all quantities necessary for the analysis and the GW signal, which is used for 
comparison.

  {\it Model s20} is the result of a simulation of the core collapse
  of a star of $20 \, \mathrm{M}_{\odot}$ of solar metallicity
  \citep{Woosley_Heger__2007__physrep__Nucleosynthesisandremnantsinmassivestarsofsolarmetallicity}
  with non-zero, but dynamically negligible, rotational velocity and
  magnetic field that was studied by~\citet{OJA2018}.  The simulation
  was performed in axisymmetry using the ALCAR/Aenus code
  \citep{Just_et_al__2015__mnras__Anewmultidimensionalenergy-dependenttwo-momenttransportcodeforneutrino-hydrodynamics}
  combining special relativistic magnetohydrodynamics, an
  approximately general relativistic gravitational potential (version `A' of the TOV
  potental of~\citet{Marek_etal__2006__AA__TOV-potential}), the SFHo 
  EOS~\citep{Steiner_et_al__2013__apj__Core-collapseSupernovaEquationsofStateBasedonNeutronStarObservations},
  and an energy-dependent, two-moment, neutrino transport scheme.  The
  core does not launch a supernova explosion.  The mass of the PNS
  increases continuously due to the ongoing mass accretion, but does
  not exceed the threshold for collapse to a black hole during the
  first second after core bounce.

{\it Model 35OC} is a 2D core-collapse simulation performed by \citet{Cerda-Duran:2013} using the general-relativistic code {\tt CoCoNuT} \citep{dimmelmeier:2002, dimmelmeier:2005}. 
The progenitor is a low-metallicity $35 {\rm M}_\odot$ star at zero-age main-sequence from~\citet{woosley:2006}. This progenitor has a high rotation rate and is usually regarded as a progenitor of long-duration gamma-ray bursts (GRBs). The simulation used the LS220 EOS of~\citet{Lattimer:1991} to describe matter at high density along with a simplified leakage scheme to approximate neutrino transport. 
This model does not produce a supernova explosion. Instead, the PNS becomes unstable to radial perturbations and collapses to a black hole 1.6 s after bounce.

\begin{figure}
\includegraphics[width=0.49\textwidth]{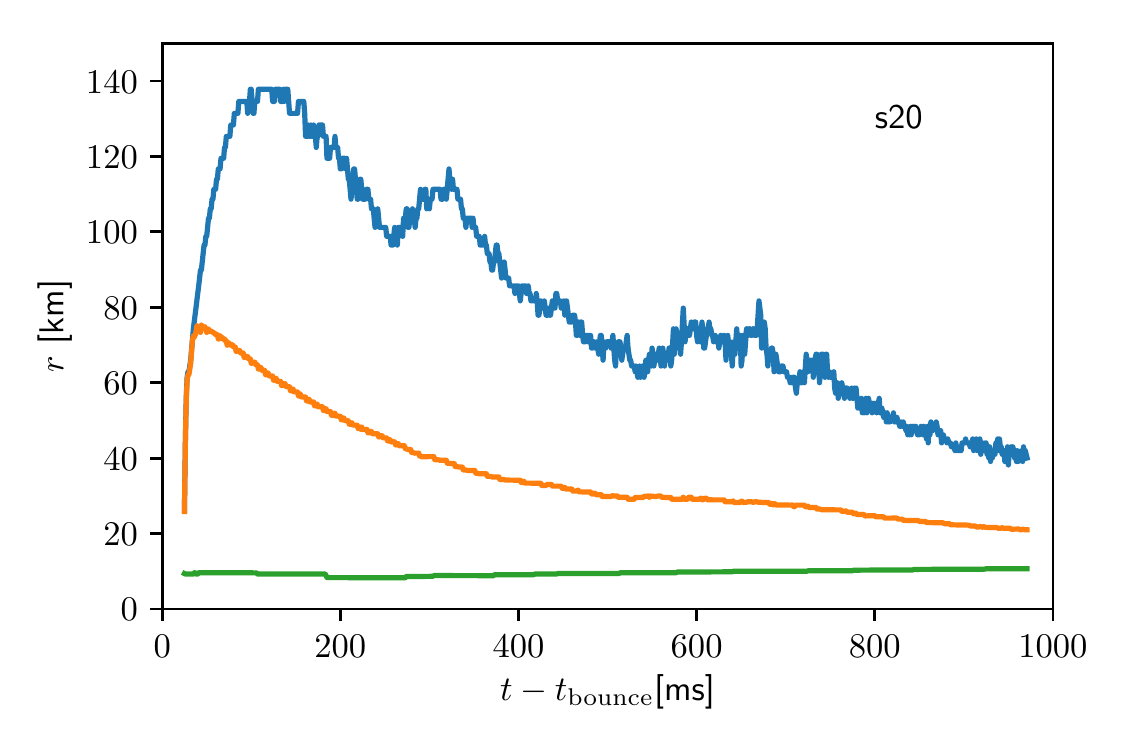}
\includegraphics[width=0.49\textwidth]{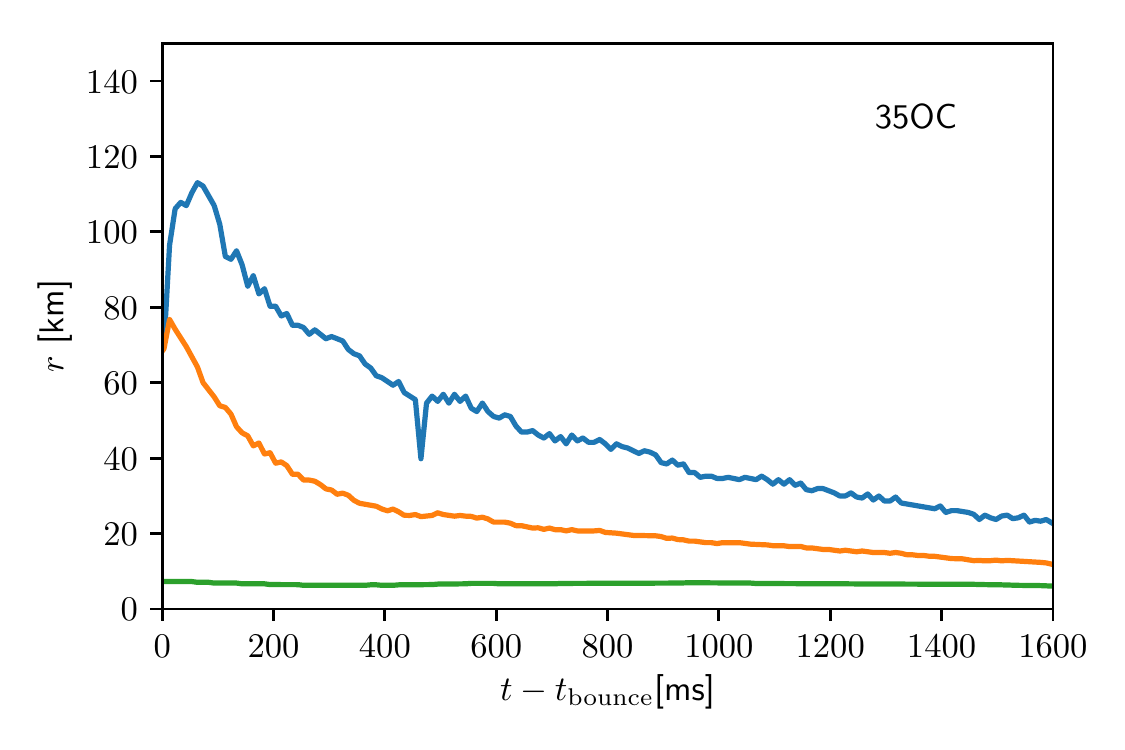}
\caption{Time evolution of the location of the shock (blue), the surface of the PNS (orange) and the inner core (green), for models $s20$ (upper panel) and $35OC$ (bottom panel). }
\label{fig:snradii}
\end{figure}

Fig.~\ref{fig:snradii} shows the time evolution of the location of the shock, the PNS surface and the outer boundary of the inner core for both simulations. We define the {\it shock} as the location where the flow becomes subsonic, the {\it PNS surface} as the radius at which $\rho=10^{11}$~g~cm$^{-3}$, and the {\it inner core} as the region where $\Gamma_1 > 2$, which is a good tracer of the region where the transition to nuclear matter has occured. 

\subsection{Numerical integration}

For the numerical calculation of the eigenmodes (``modes" hereafter) we follow a similar procedure as in \cite{Torres-Forne:2018}, i.e.
we solve the perturbation equations for different values of $\sigma$, and then search for the values of $\sigma$ with vanishing radial 
displacement at the shock ($\eta_r|_{\rm shock}$ =0), by means of a bisection algorithm. The perturbation equations differ depending
if $l=0$ or $l\ne0$, but in both cases the system can be cast as in Eq.~(\ref{eq:compact}). We integrate this system of six coupled
ODEs outwards from the center to the shock radius. The integration is performed 
using a second-order implicit method (trapezoidal rule). Numerical tests for this procedure can be found in~\cite{Torres-Forne:2018}. 

Given that we are integrating a system of six ODEs, we have to provide six boundary conditions. Since we are using a staggered 
grid, our first integration point is not at $r=0$. This implies that we have to provide non-zero values for all quantities at this point
to perform the integration. Due to the regularity conditions seen in Section~\ref{sec:bcs}, it is sufficient to fix the value of three quantities,
$\eta_r$, $\delta \hat Q$ and $\delta \hat \psi$, and the remaining three are automatically known. 

The value of  $\eta_r$ can be set arbitrarily and it fixes the amplitude  of the eigenfunction. The two additional variables 
have to be fixed such that the boundary conditions at the shock location, Eqs.~(\ref{eq:BCS1}) and (\ref{eq:BCS2}), are fulfilled.
To ensure this, we use a shooting method, which consists in varying the values
 of $\delta \hat Q$ and $\delta \hat \psi$ at the innermost radial point, integrating outwards, and checking whether the boundary conditions are fulfilled. 
 We perform this iteration using a vectorial Newton-Raphson method, where the derivatives of the Jacobian are computed numerically using a stencil
 of the size given by the previous step. 
 
We also compute the eigenmodes using an alternative numerical method. The description of this method and the comparison with the
first one can be found in Appendix~\ref{sec:appendix:alt}. In all tested cases, the differences in the eigenfrequencies between both methods 
were smaller than $0.1\%$. In some cases (for very low frequencies) the alternative method shows numerical convergence problems that are not found
in the first method. Therefore, all results presented in this paper have been obtained using the first method, which is more robust.

The computation of $\mathcal G$, defined in Eq.~(\ref{eq:G}), involves some degree of arbitrariness because it can be computed either from
the gradient of the pressure (${\mathcal G}_P \equiv \partial_r P / \rho h$) or from the gradient of the lapse (${\mathcal G}_\alpha \equiv -\partial_r \ln \alpha$). 
Unless stated otherwise  we use $\mathcal G = \mathcal G_P$. We explore the 
effect of the definition of $\mathcal G$ in the eigenmode calculation in Section~\ref{sec:spec}.

\subsection{Eigenmode classification}

The procedure we have just described allows us to compute the eigenvalues, along with their corresponding eigenfunctions, 
for a set of time slices of a simulation. However, it does not provide information about the nature of each eigenmode. Stellar oscillations
can be classified according to the dominant restoring force giving rise to them, either pressure ($p$-modes) or buoyancy ($g$-modes). 
The local quantities determining the character of the modes are the Lamb frequency, $\mathcal L$, and the Brunt-V\"ais\"al\"a frequency, 
$\mathcal N$. Pressure supported (sound) waves with frequency $\sigma$ are possible in regions of the star in which $\sigma^2 > \mathcal L^2, \mathcal N^2$, 
while buoyancy supported (gravity) waves are possible in regions with  $\sigma^2 < \mathcal L^2 \mathcal, N^2$ \citep[see e.g.][]{Cox:1980}. The regions
of the star where $\sigma^2$ is between $\mathcal L^2$ and $\mathcal N^2$ are evanescent  and no waves propagate in this region. 
Note that if $\mathcal N^2 <0$ gravity waves are not possible because the system is convectively unstable.
Using these properties several classification procedures are possible, which have been developed in the context of asteroseismology \citep[see e.g.][]{Unno:1979,Cox:1980}.

\begin{figure}
\includegraphics[width=0.47\textwidth]{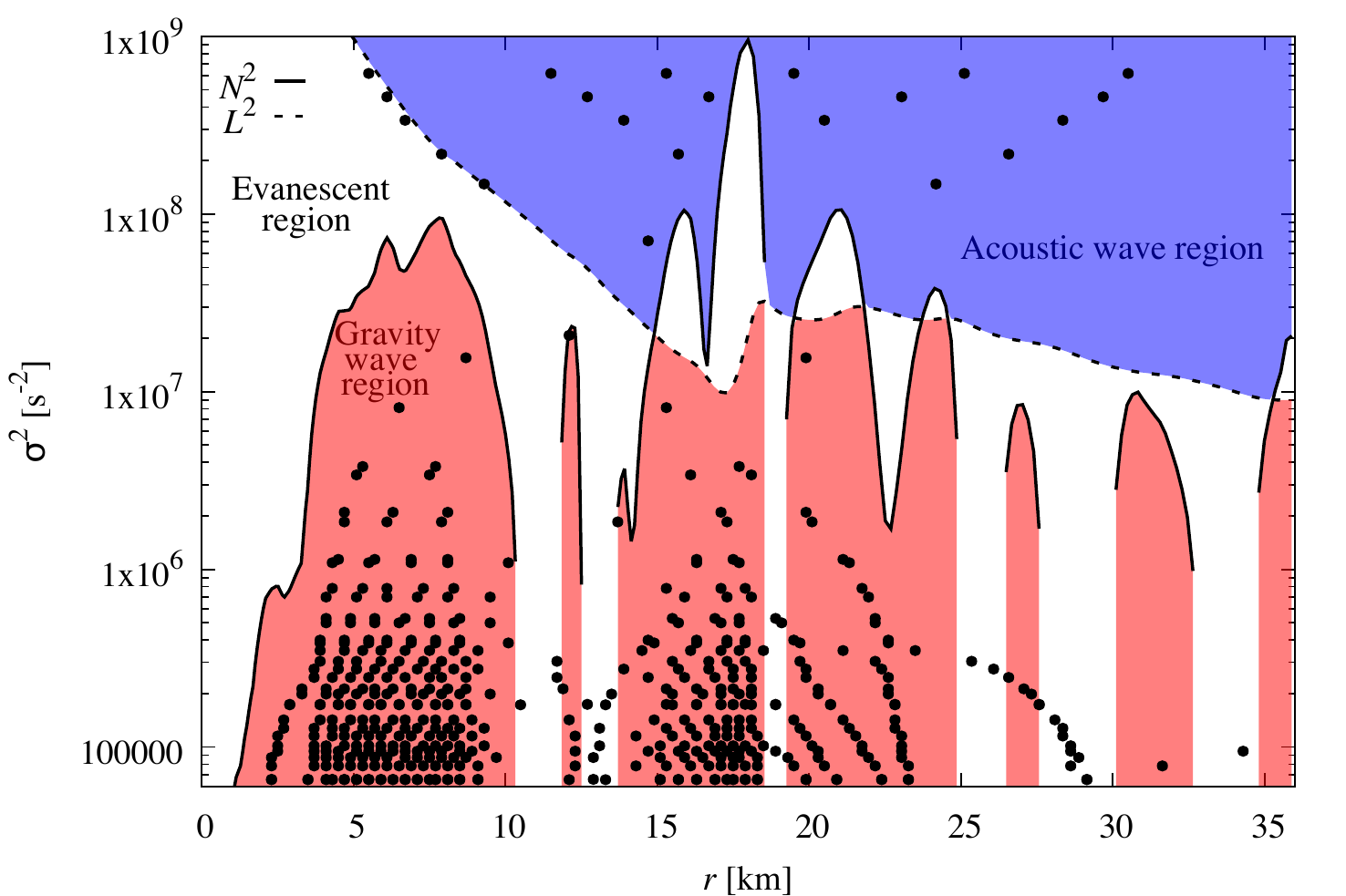}\\
\includegraphics[width=0.23\textwidth]{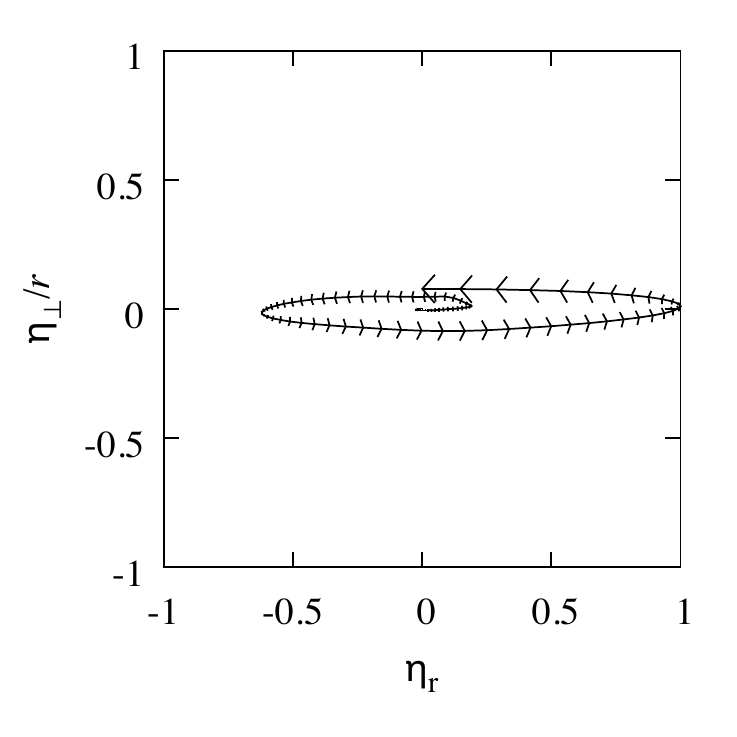}
\includegraphics[width=0.23\textwidth]{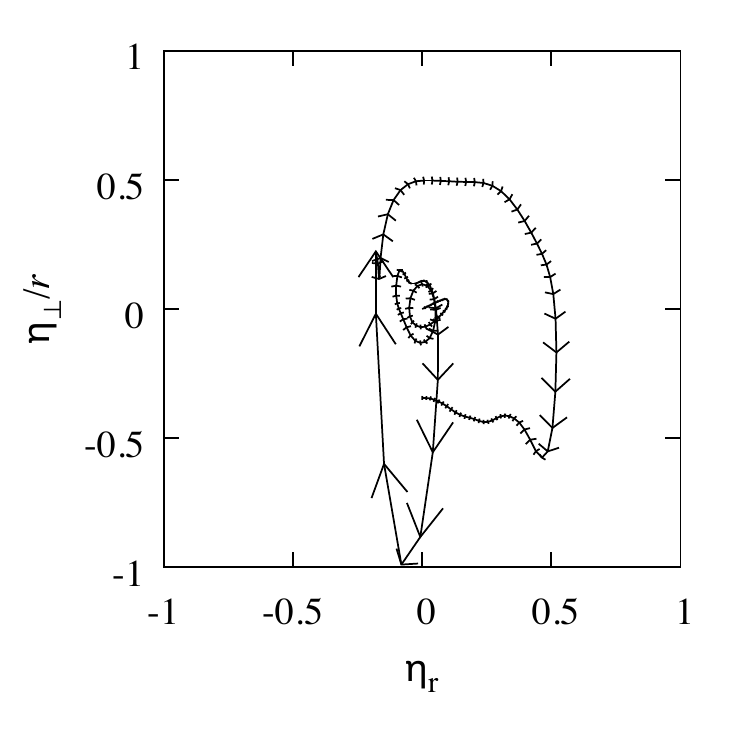}
\caption{Propagation and phase diagrams for $l=2$ modes in model 35OC at $1.3$~s post bounce. {\it Upper panel}: Propagation diagram showing the radial profile of the Brunt-V\"ais\"al\"a frequency ($\mathcal N^2$, solid black line) and the Lamb frequency ($\mathcal L^2$, dashed black line). Note that in some regions $\mathcal N^2<0$ and no gravity waves are possible. Colours indicate the acoustic wave region (blue, $\sigma^2 > \mathcal N^2, \mathcal L^2$), the
gravity wave region (red, $\sigma^2 < \mathcal N^2, \mathcal L^2$), and the evanescent region (white). Black dots indicate the location of the radial nodes of $\eta_r$ for all the eigenmodes found at different $\sigma$. {\it Lower panels}: phase diagram for  two eigenmodes at $3542$~Hz ($\sigma^2 = 5.0 \times 10^8$~s$^{-2}$) corresponding to a p-mode (lower left) and at $237.8$~Hz ($\sigma^2=2.2 \times 10^6$~s$^{-2}$) corresponding to a g-mode (lower right).
Arrows indicate the direction of increasing $r$. The trajectory rotates clockwise in gravity wave regions and counter-clockwise un sound wave regions. 
}
\label{fig:diagrams}
\end{figure}

\begin{figure*}
\includegraphics[width=0.45\textwidth]{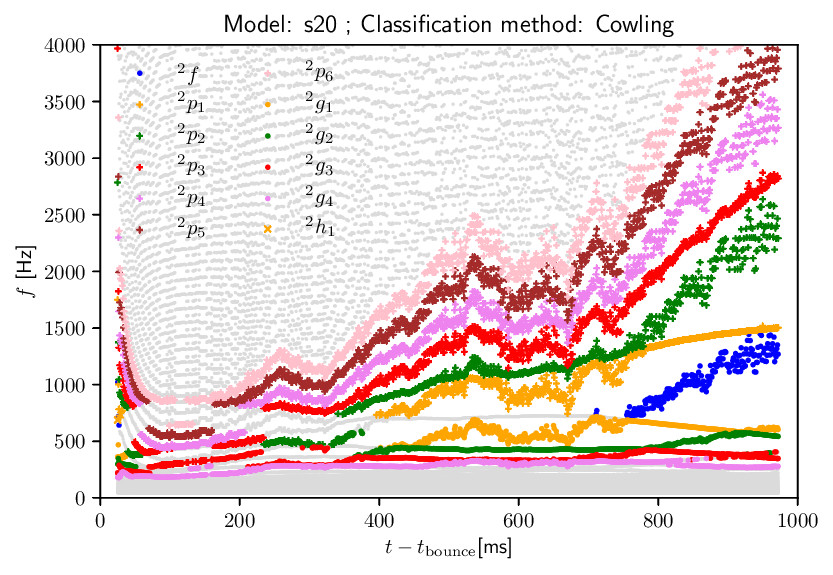}
\includegraphics[width=0.45\textwidth]{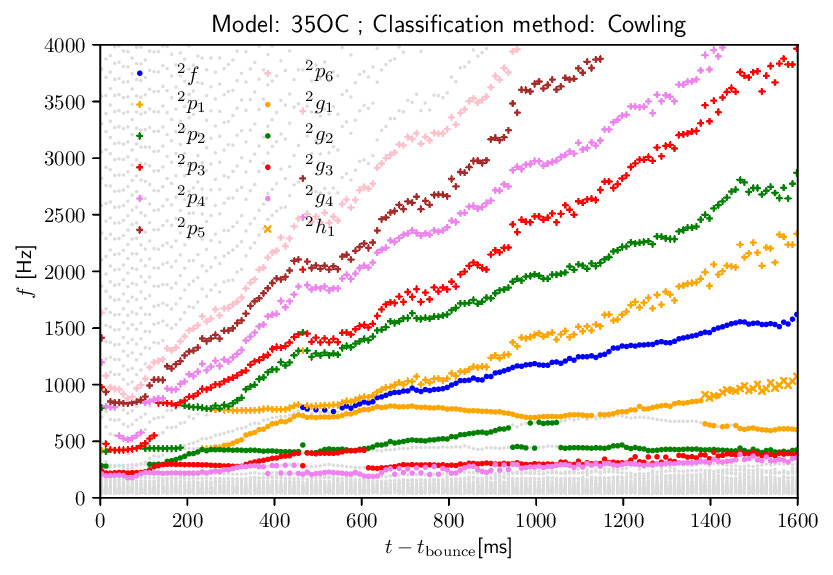}\\
\includegraphics[width=0.45\textwidth]{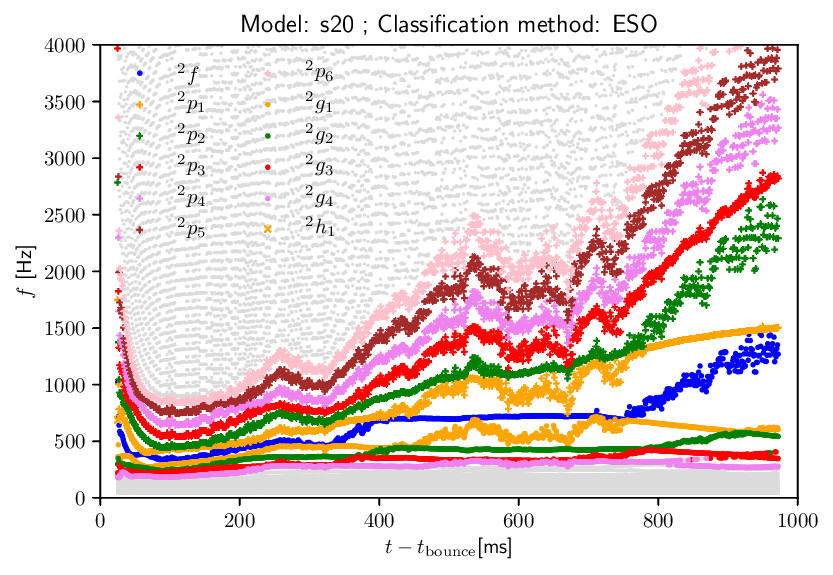}
\includegraphics[width=0.45\textwidth]{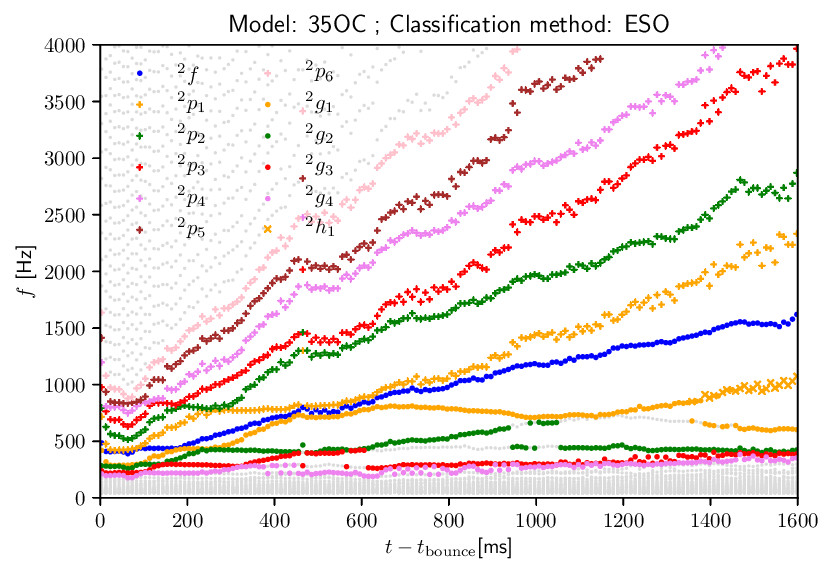}\\
\includegraphics[width=0.45\textwidth]{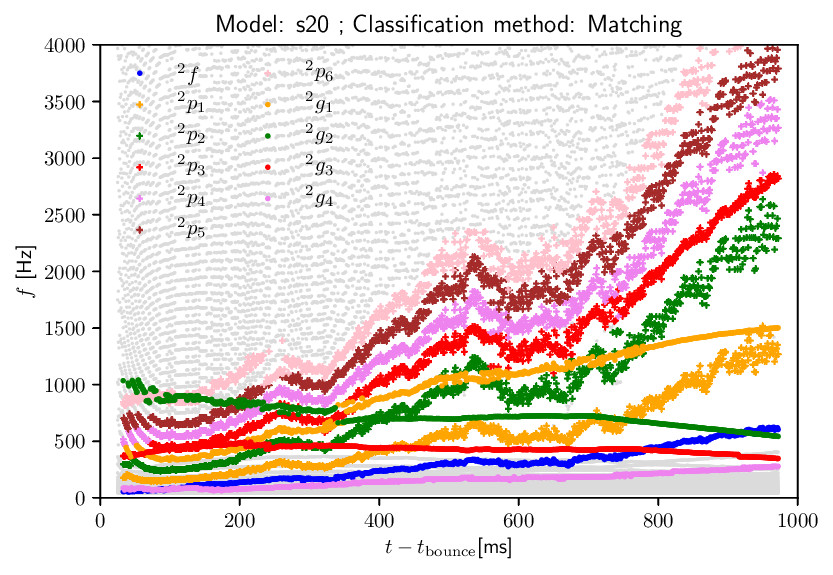}
\includegraphics[width=0.45\textwidth]{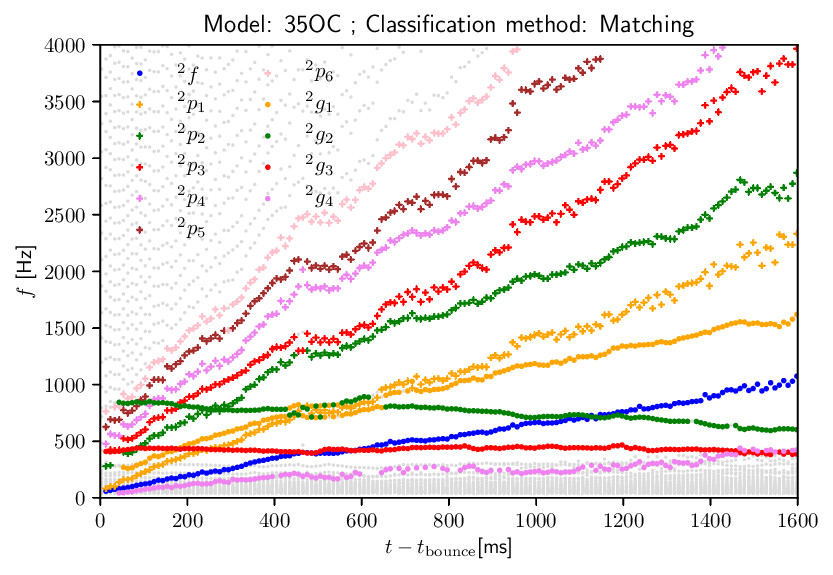}\\
\includegraphics[width=0.45\textwidth]{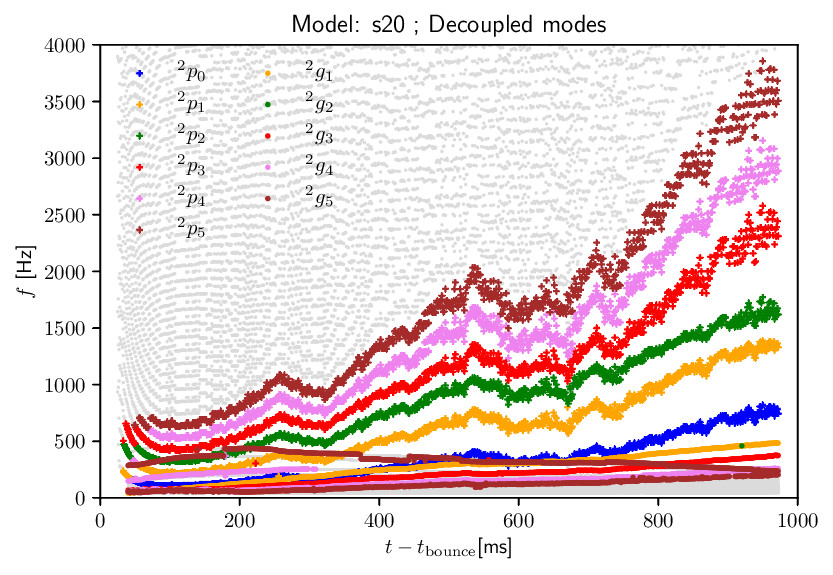}
\includegraphics[width=0.45\textwidth]{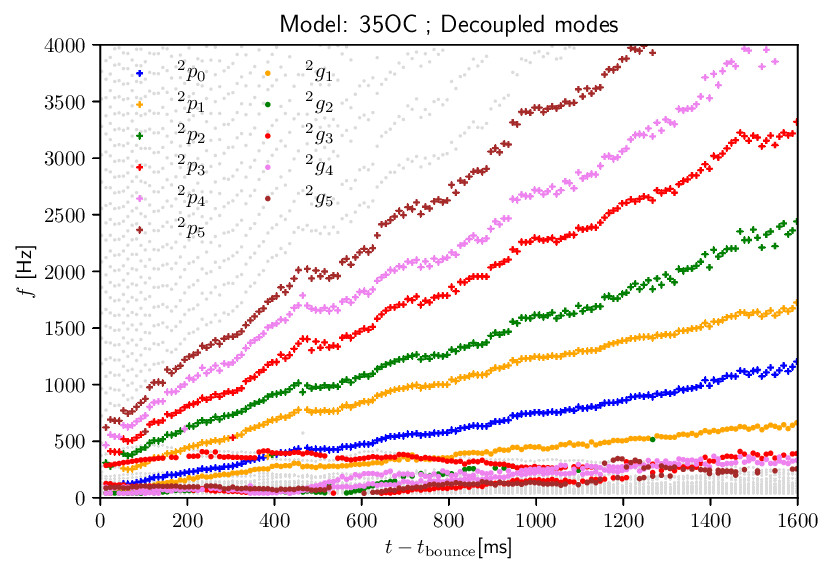}
\caption{Evolution of the eigenfrequencies for $l=2$ for models s20 (left panels) and 35OC (right panels)
using the Cowling (upper panels), ESO (second row) and matching (third row) classification procedures.  The bottom
panels correspond to the decoupled computation of the modes. A selection of the classified modes (indicated in the legends) is plotted in colours and the rest of the modes are plotted in grey. }
\label{fig:modes}
\end{figure*}

\begin{figure*}
\includegraphics[width=0.99\textwidth]{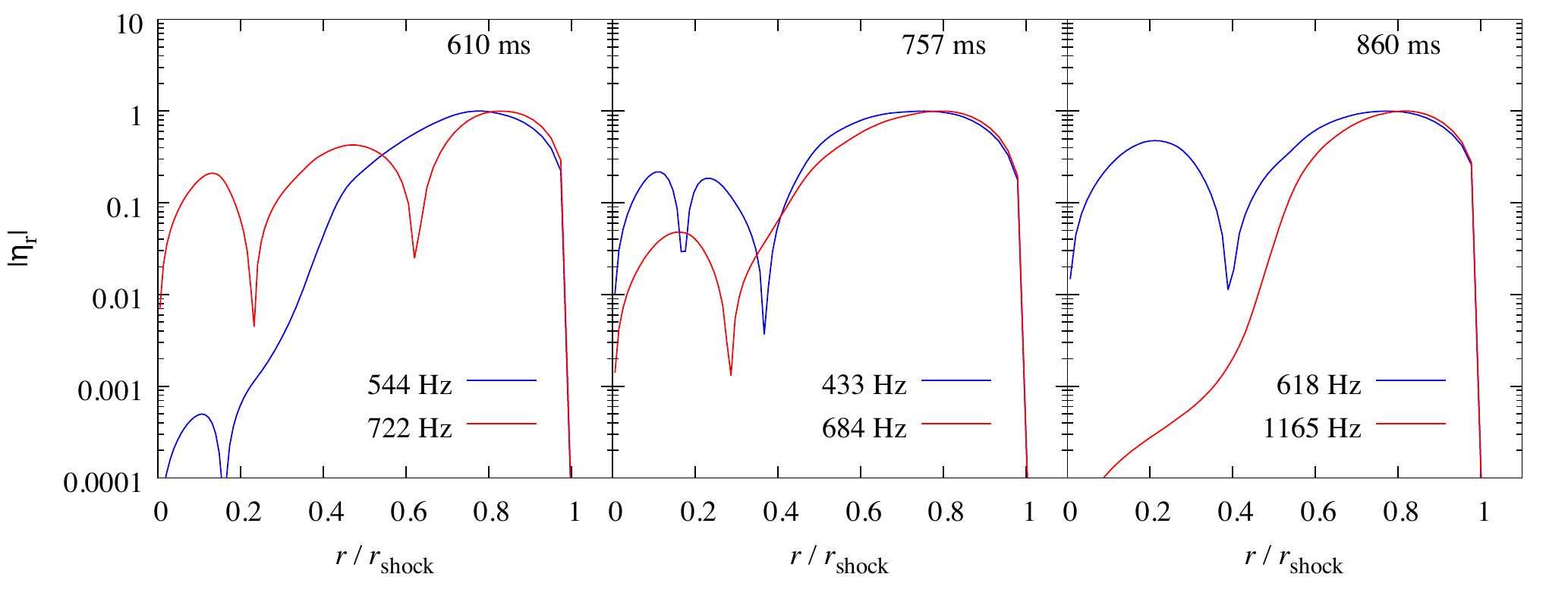}
\caption{Amplitude of the radial eigenfunction for the fundamental $l=2$ $^2 f$ mode (blue) and the $^2 g_1$ g-mode (red) for the s20 model, classified according to 
the ESO scheme, around the time of the avoided crossing ($t-t_{\rm bounce}\sim 750$~ms).}
\label{fig:crossing}
\end{figure*}

\subsubsection{Cowling classification}

The first classification of non-radial oscillation modes of spherical stars was introduced by \cite{Cowling:1941}\footnote{We warn the 
reader not to confuse the Cowling approximation (static space-time approximation) and the Cowling classification procedure of modes, both introduced
in \cite{Cowling:1941}.}. For stars with monotonically decreasing $\mathcal L^2$ (note that it is proportional to $r^{-2}$) and monotonically increasing $\mathcal N^2$ \citep[typical of simple stratified equilibrium models; see e.g.][]{Cox:1980} there is a critical frequency above which only sound waves can propagate and below which only gravity waves are possible. This allows for a very simple classification purely based in the number of nodes of the radial part of the eigenfunction, $\eta_r$. The mode with zero radial nodes is the fundamental mode or f-mode, denoted as $^lf$. Modes with higher frequencies are p-modes, denoted as $^lp_n$, with increasing number of nodes, $n$, for increasing frequency. In a similar way, g-modes, denoted as $^lg_n$, have frequencies lower than the f-mode and have increasing number of nodes, $n$, for decreasing frequency. A variant of this classification has been used in \cite{Torres-Forne:2018}. In that work, h-modes (hybrid) were introduced to distinguish modes above or below the f-mode, but with the same number of nodes. The upper panels of Fig.~\ref{fig:modes} below show the eigenmodes classified according to the Cowling classification (variant described in  \cite{Torres-Forne:2018}).

Although it can serve as a guide, the Cowling classification does not work
properly in every case, in particular in those cases with no critical frequency. In those cases, it may happen that, for a given frequency, there are regions of the
star supporting gravity waves at the same time as other regions support sound waves, sometimes separated by evanescent regions (see upper panel of 
Fig.~\ref{fig:diagrams}). In those cases the ordering devised by Cowling may not apply and one has to rely upon a more general procedure \citep[see][for a deeper discussion]{Unno:1979,Cox:1980}. An indication that this is indeed a problem in our models is the necessity of introducing h-modes in the classification. As we 
show below, this is an artefact of the classification scheme.

\subsubsection{ESO classification} 

\begin{figure*}
\includegraphics[width=0.48\textwidth]{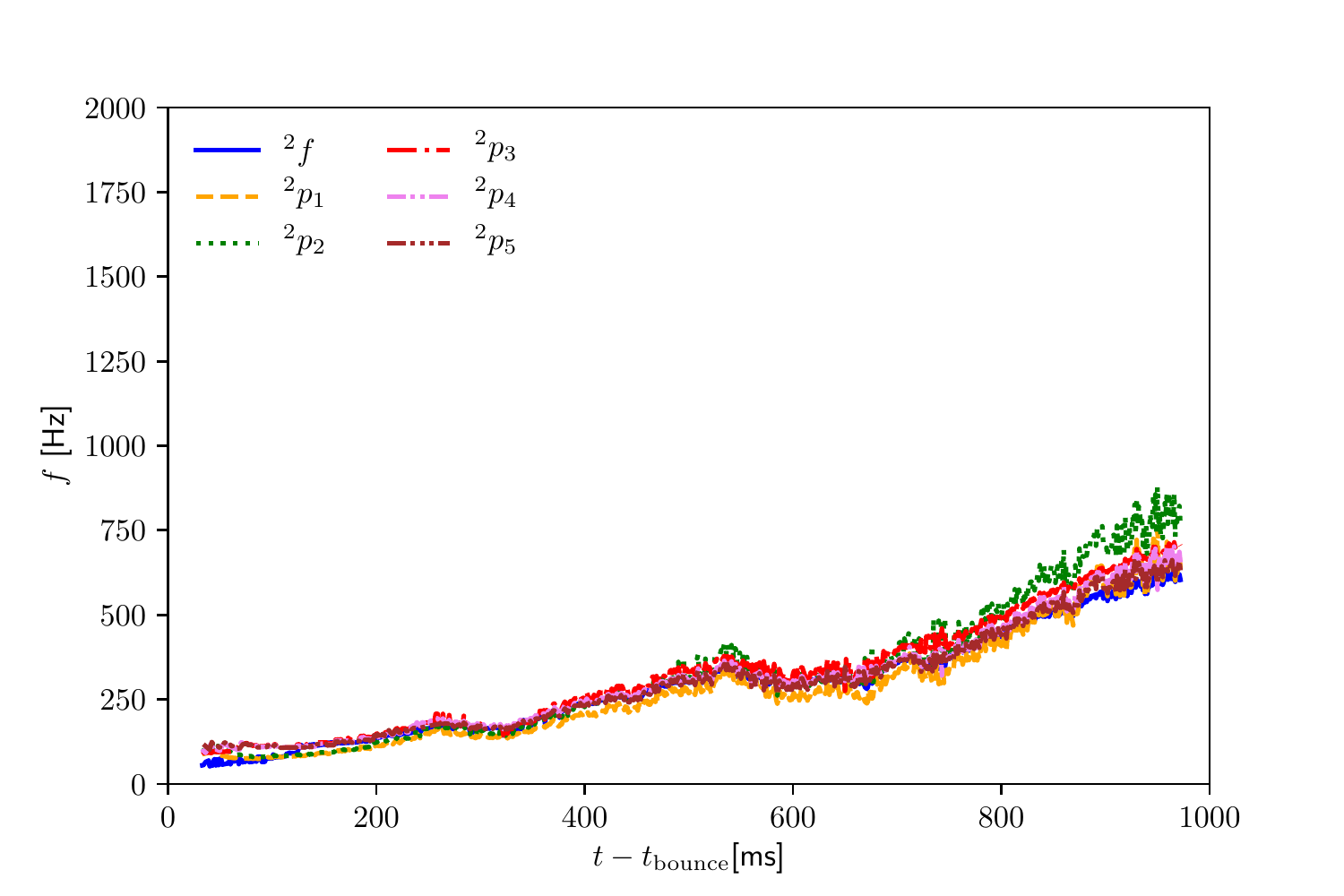}
\includegraphics[width=0.48\textwidth]{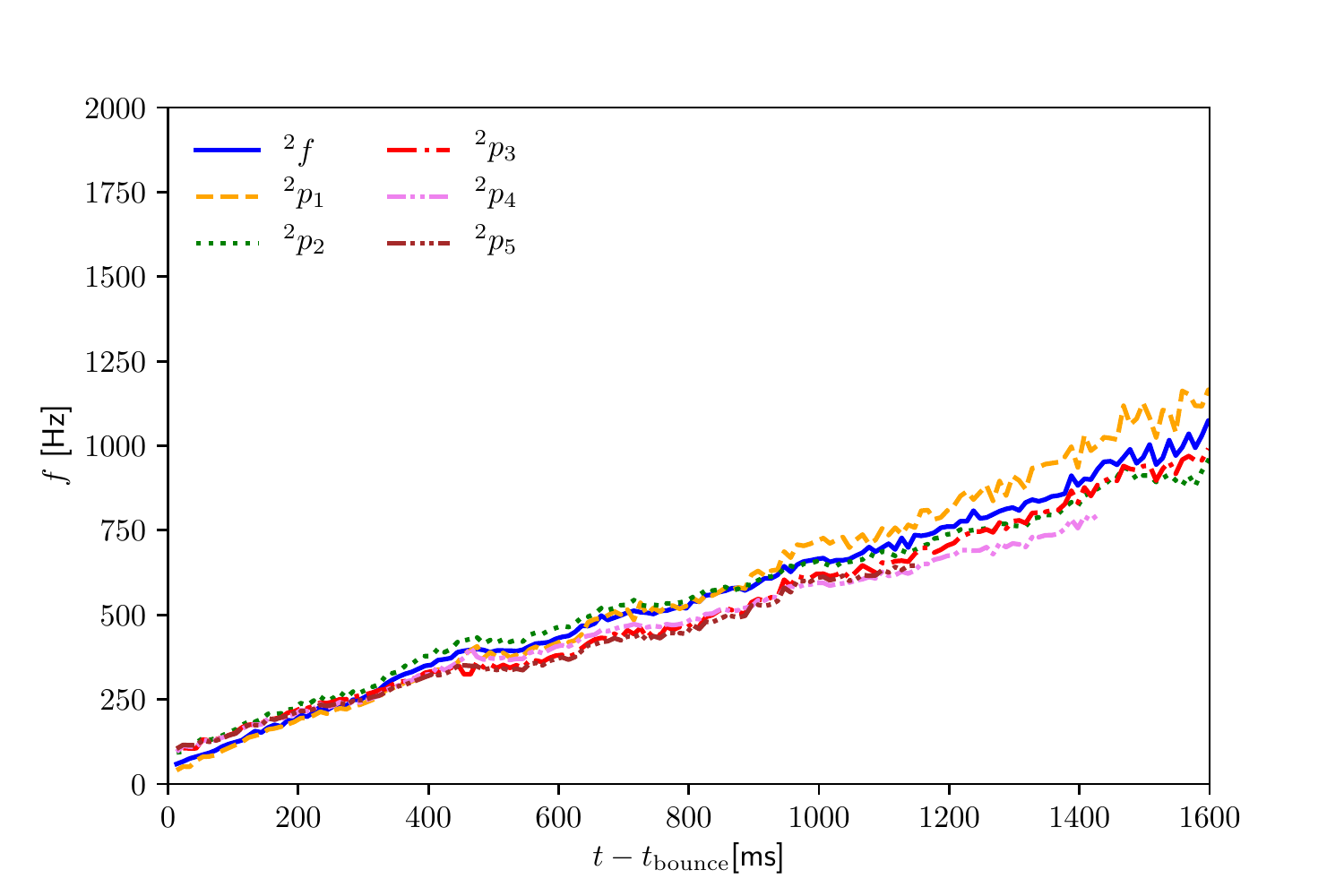}
\caption{Evolution of the frequencies of the $^2p_n$ modes divided by $n+1$ compared to the $^2f$ mode, both for the s20 model (left) and the 35OC model (right).}
\label{fig:fscaled}
\end{figure*}

To overcome the problems of the Cowling classification \cite{Eckart:1960}, \cite{Scuflaire:1974} and \cite{Osaki:1975} developed a classification
scheme (ESO scheme, hereafter) based not only on the number of radial nodes, but also in the character of each node. Using the radius $r$ as a parameter,
each mode can be plotted in the  $\eta_r$ vs $\eta_\perp$ phase diagram. Nodes in the radial direction correspond to crossings of the ``$x$" axis ($\eta_r$).
For g-modes, the trajectory of the mode in the phase diagram (parametrised with $r$) is clockwise (see lower right panel of Fig.~\ref{fig:diagrams}, for an example), 
while p-modes have counter-clockwise trajectories (lower left panel of Fig.~\ref{fig:diagrams}). For frequencies in regions supporting gravity waves and regions supporting sound waves, the trajectory in the phase diagram is clockwise and counter-clockwise in each of the regions. The ESO scheme is based on the number of clockwise turns minus the number of counter-clockwise turns in the phase diagram, $n_{\rm ESO}$. For $n_{\rm ESO} > 0$ the mode is predominantly a g-mode and it is classified as a $^lg_{n}$, with $n=n_{\rm ESO}$. For $n_{\rm ESO}<0$ the mode is predominantly a p-mode and it is classified as a $^lp_n$ with $n = -n_{\rm ESO}$. The mode with $n_{\rm ESO}=0$ is the f-mode. The second row of panels of Fig~\ref{fig:modes} shows the ESO classification method applied to our models.

The ESO scheme significantly improves the classification of the modes, although it still has some drawbacks. One of the problems is the existence of trapped modes. If one looks at the propagation diagram in the upper panel of Fig.~\ref{fig:diagrams}, at mid-range frequencies ($\sigma^2\sim 5\times 10^8$~s$^{-2}$) there is a region within $10$~km from the center where only gravity waves are possible and at the same time sound waves are only possible above $\sim 12$~km. Both regions are disconnected by an evanescent region, so in principle it is possible to have trapped modes inside each of the two regions, corresponding to a g-mode and a p-mode, respectively, at the same (or similar) frequencies. In practice these two modes interact with each other channeling through the evanescent region and giving raise to a more complex mode which is a hybridisation of both. This effect can be observed in both of our models, when looking at the time evolution of the eigenfrequencies (Fig.~\ref{fig:modes}, see ESO classification). As the frequency of any two modes becomes similar the phenomenon of the avoided crossing appears.

To illustrate the phenomenon of the avoided crossing, let us consider the $^2 f$ and $^2 g_1$ modes at $\sim750$~ms in the model s20. According to the 
ESO classification scheme, the $^2f$ mode has higher frequency than the $^2 g_1$ during the avoided crossing (see left panel in the second row of Fig~\ref{fig:modes}). This produces an abrupt change of frequency near $750$~ms, which may appear as an artefact of the classification scheme (it would look more natural if they crossed). Fig.~\ref{fig:crossing} shows $\eta_r$ for these two modes before ($610$~ms), during ($757$~ms) and after ($860$~ms) the crossing. Before the crossing (left panel) the $^2f$ mode appears more concentrated in the outer parts of the system (outside the PNS), while the $^2g_1$ mode extends to the interior of the PNS. However, after the crossing (right panel) the situation is reversed, appearing as if the ESO scheme would have misclassified the modes. During the avoided crossing, both modes hybridise (middle panel) and are actually quite similar (except for the number of nodes), which is the result of the process described in the previous paragraph. This phenomenon is well known in asteroseismology, and not a numerical artefact of the ESO scheme. However, the ESO classification scheme  poses a problem for the purpose of this work. Our goal is to learn how the eigenmodes behave during the post-bounce evolution to try to devise ways, in the future, to infer properties of the PNS based on GW observations. For this purpose it is crucial to characterize the GW features seen in spectrograms (mostly described as raising archs) and classify together modes with similar features (e.g. localised in the same part of the PNS) which are likely to be excited with similar energy and produce a similar GW output, during the post-bounce evolution. For this reason we need a method that is based in the similarity of the eigenfunctions and not in the number of nodes.

\subsubsection{Matching classification} 
\label{sec:match:class}

In this work, we present a new classification procedure which is not based in the number of nodes but in the shape of the eigenfunctions. Our procedure traces the eigenmodes in time by finding the best match between the shape of the eigenfunction in each time step and those from the previous time steps. We have found that this procedure works best when the matching is done backwards in time. The details of the matching algorithm can be found in  Appendix~\ref{sec:appendix:class}. Hereafter, we refer to this algorithm as the {\it matching classification}.  Note that the algorithm does not give a proper classification in the sense that it does not tell g-modes from p-modes or the f-mode, it just groups modes at different times in groups according to similarity. To tag each mode sequence with an appropriate class we use as a reference the ESO classification at the last time available, with some additional modifications that we discuss next. The third row panel of Fig.~\ref{fig:modes} shows the classification results for the matching scheme. With the new procedure the behaviour in time of the eigenmodes is smoother and there are mode crossings at places where avoided crossings appeared with the previous two classification methods. In particular, in the crossing of modes in the s20 model discussed above, the shapes are preserved (modes are swapped with respect to Fig.~\ref{fig:crossing}), but the number of nodes changes. This feature is consistent for all modes and a more detailed analysis of the eigenmode shape is given in the next section. Note that in the new classification scheme, the number of nodes indicated in the name of the class (e.g. $^2 g_1$, one node) is not indicative of the actual number of nodes. This new mode classification method is not perfect, and some modes are still clearly misclassified (specially high order g-modes). However, for low order modes it gives consistent results. We will argue in the next sections that these modes are the most relevant for GW emission, and therefore our classification should be sufficient for this purpose.

To better understand the classification of the modes we solve the eigenvalue problem decoupling p-modes and g-modes. This can be achieved by 
setting $\mathcal B=0$ or $c_s^2 \to \infty$, respectively, as described in \cite{Torres-Forne:2018}.  The two lower panels of Fig.~\ref{fig:modes} shows
the decoupled modes in these limits.  We use this information to retag some of the modes classified by the matching algorithm to better match the identification made in the decoupled computation. This result also confirms that the avoided crossings seen in the Cowling and ESO classifications are related to crossings of the decoupled modes and therefore our matching algorithm is unveiling these crossings properly. We note that many of the decoupled modes also differ  significantly in frequency with respect to the corresponding modes computed with the full system. In particular, g-modes tend to have higher frequencies in the full system, likely due to the presence of acoustic wave regions where those waves can propagate significantly faster that gravity waves. 

One of the consequences of the matching classification scheme is that it renders unnecessary to introduce h-modes to classify all modes. For example, what was misclassified as an $^2h_1$ mode in the Cowling and ESO classification for the 35OC model, as well as in \cite{Torres-Forne:2018}, is actually classified as the f-mode with the new scheme and all p-modes have a value of $n$ displaced in one unit with respect to our previous classifications. A more definitive proof that our matching  procedure  classifies correctly the f-mode and the p-modes is that the frequency of the $^2p_n$ modes is approximately an integer number ($n+1$) the frequency of the f-mode (See Fig.~\ref{fig:fscaled}). This relation is significantly better than the one found in \cite{Torres-Forne:2018}, which misclassified the f-mode.  It also clarifies the intriguing feature found in \cite{Torres-Forne:2018}, of a h-mode with a frequency which was an integer fraction of the higher order p-modes. This mode simply was the f-mode.

Finally, we would like to indicate that there are other classification schemes in the literature
\citep[e.g.][]{Takata:2012}, which tried to overcome the limitations of the ESO scheme using different
approaches. How these classification schemes compare to our approach is something that could be
explored in future work.

 \begin{figure*}
\includegraphics[width=0.43\textwidth]{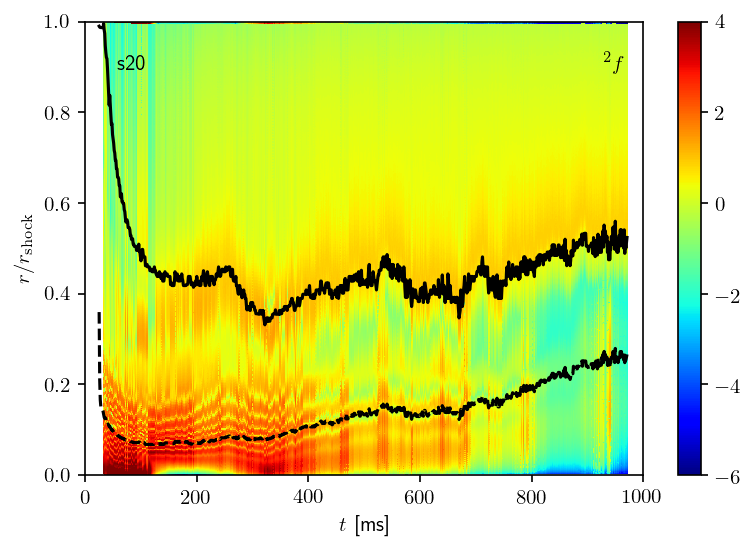}
\includegraphics[width=0.43\textwidth]{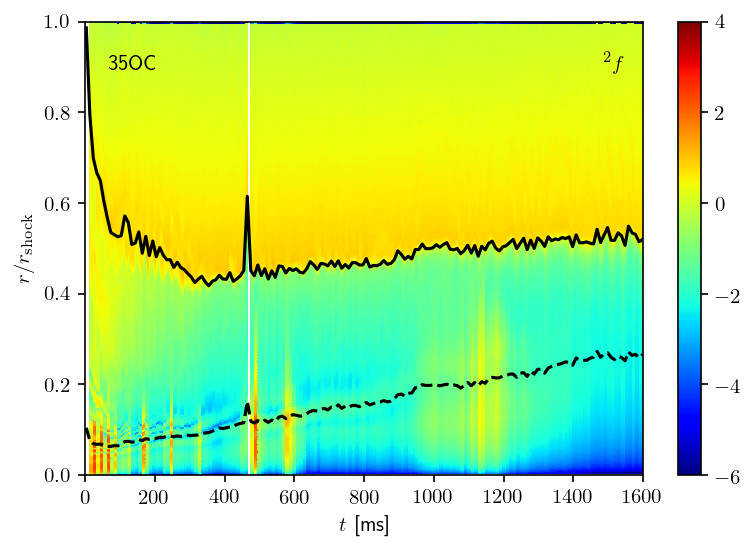}\\
\includegraphics[width=0.43\textwidth]{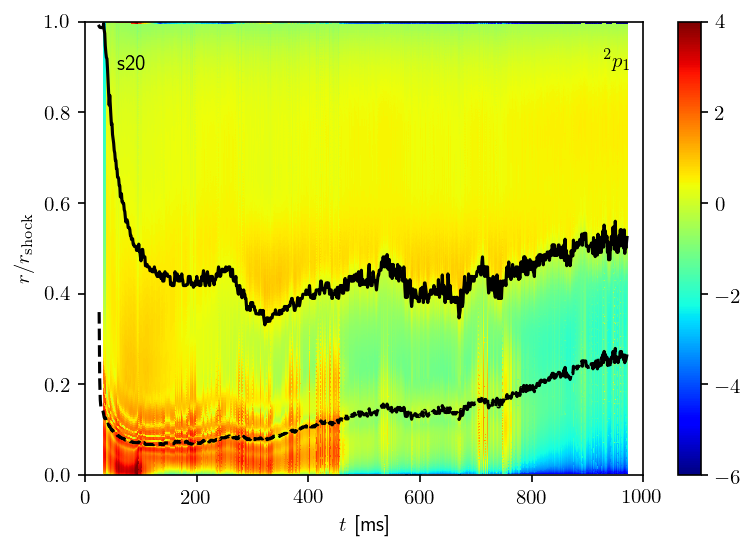}
\includegraphics[width=0.43\textwidth]{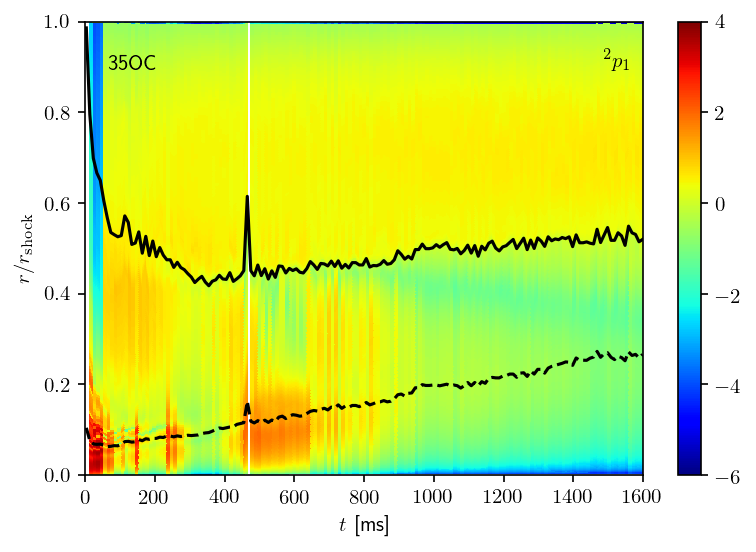}\\
\includegraphics[width=0.43\textwidth]{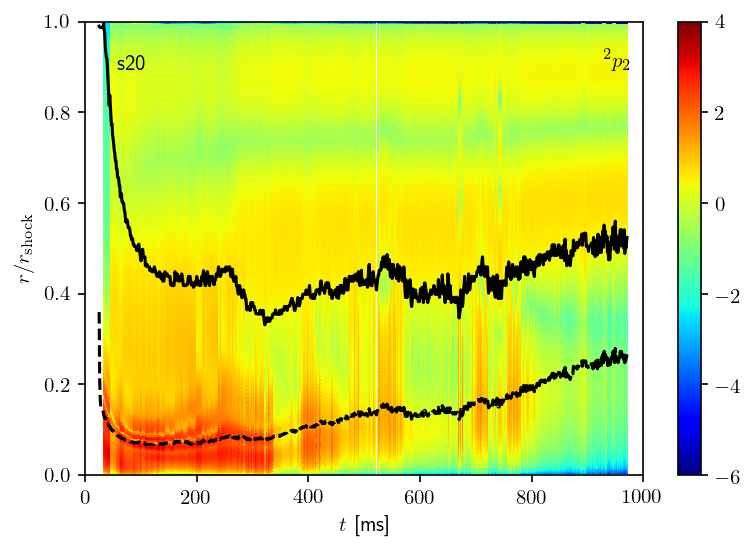}
\includegraphics[width=0.43\textwidth]{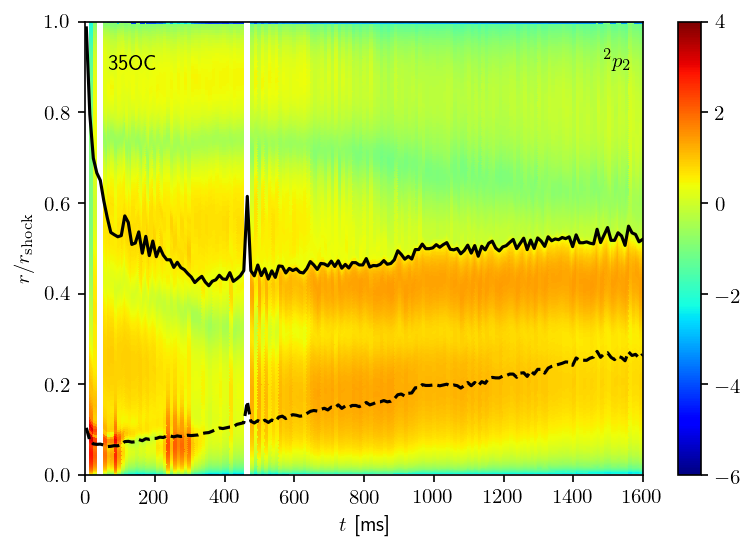}\\
\includegraphics[width=0.43\textwidth]{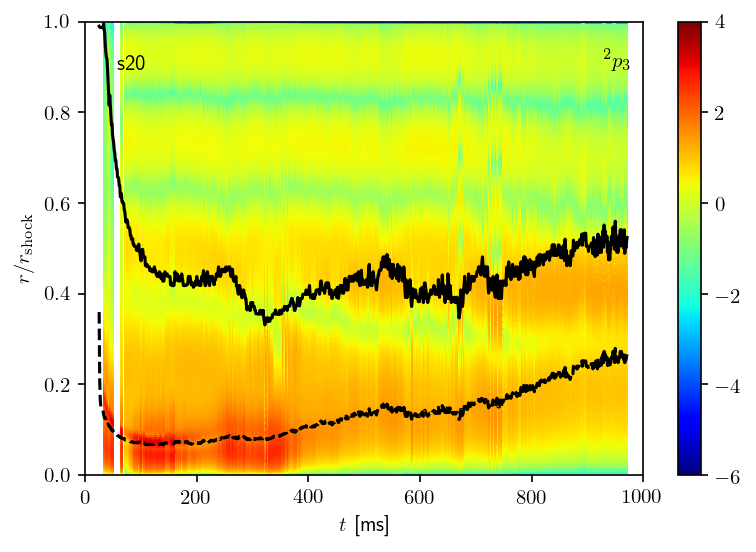}
\includegraphics[width=0.43\textwidth]{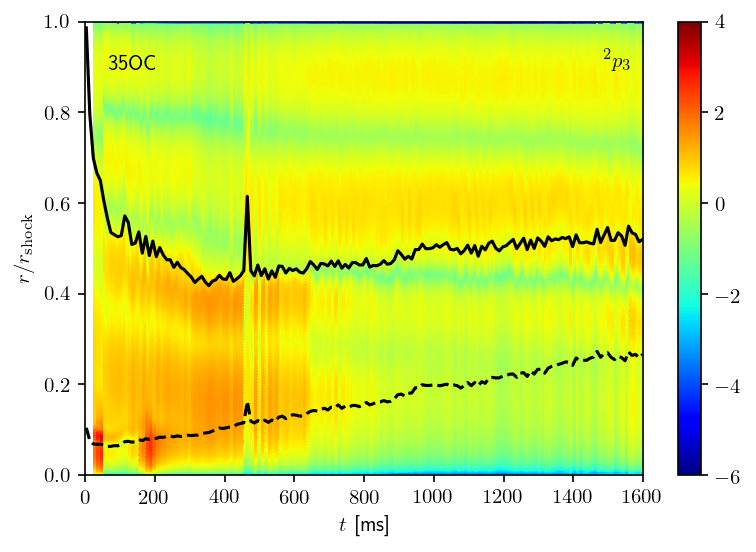}
\caption{Time evolution of the radial profile of the logarithm of the energy density, $\mathcal{E}(r)$, for a selection of modes classified as p-modes and the f-mode for models s20 (left panels) and 35OC (right panels). The solid black line indicates the position of the PNS surface and the dashed black line the surface of the inner core. The radius is normalised to the shock location. White stripes indicate times in which no modes were found.}
\label{fig:morpho_p}
\end{figure*}

\begin{figure*}
\includegraphics[width=0.43\textwidth]{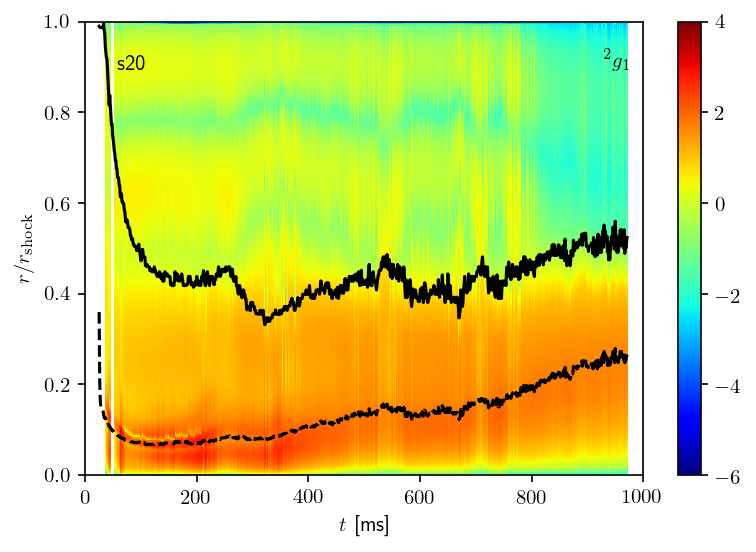}
\includegraphics[width=0.43\textwidth]{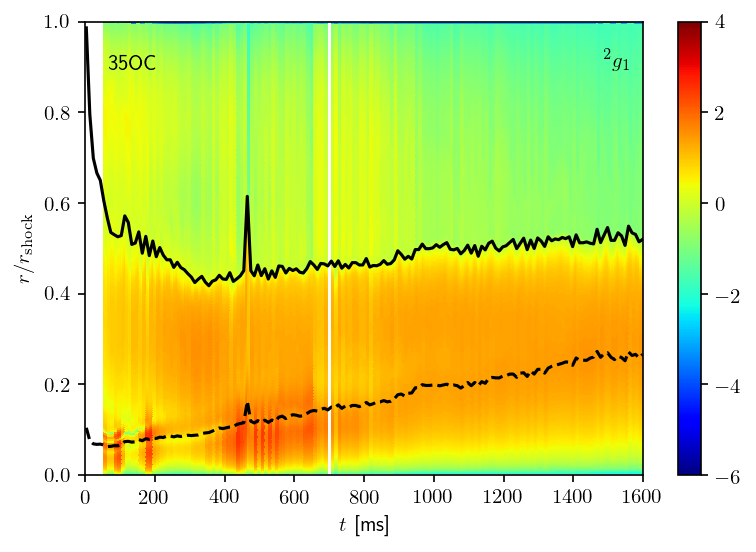}\\ 
\includegraphics[width=0.43\textwidth]{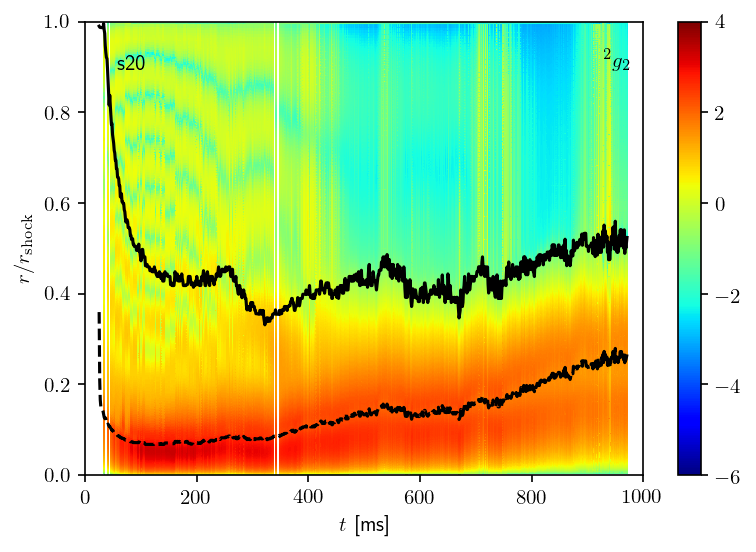}
\includegraphics[width=0.43\textwidth]{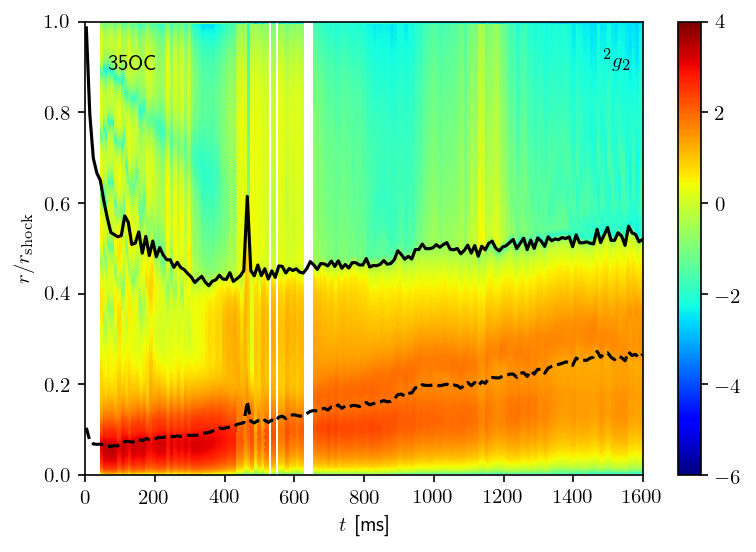}\\
\includegraphics[width=0.43\textwidth]{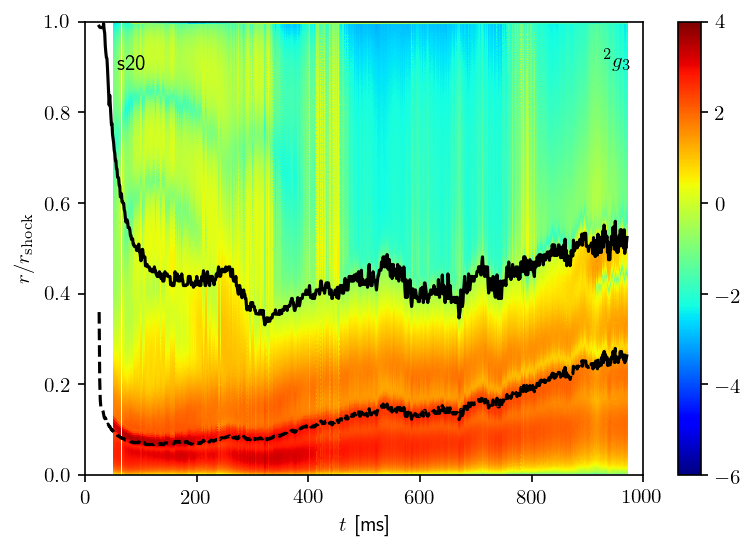}
\includegraphics[width=0.43\textwidth]{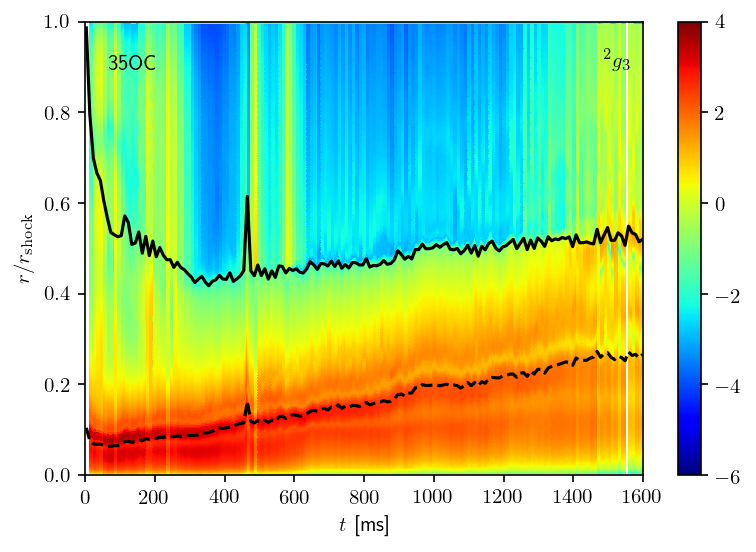}
\caption{Same as Fig.~\ref{fig:morpho_p}, but for a selection of modes classified as core g-modes.}
\label{fig:morpho_gcore}
\end{figure*}

\begin{figure*}
\includegraphics[width=0.43\textwidth]{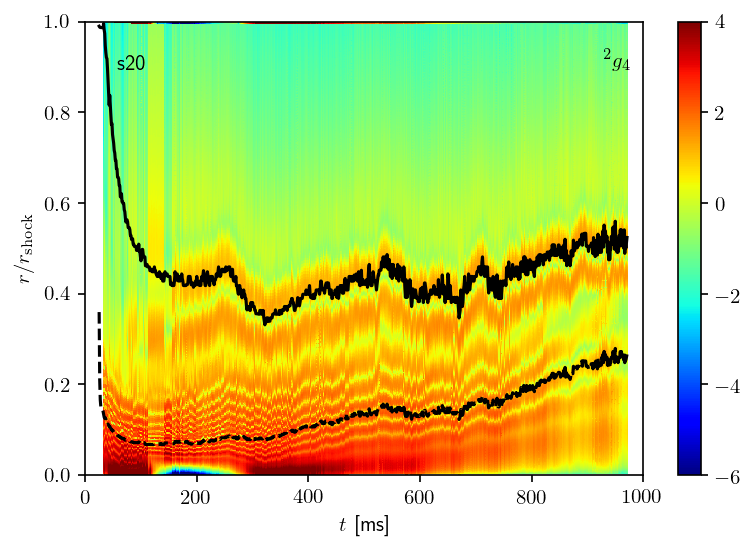}
\includegraphics[width=0.43\textwidth]{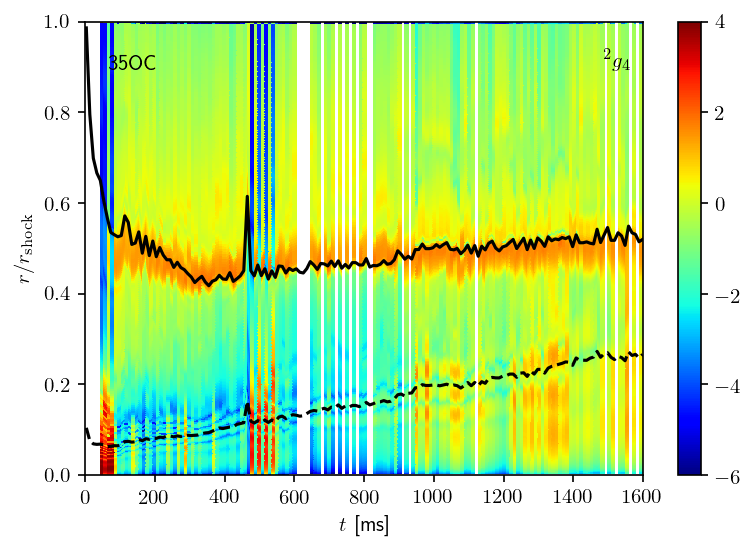}\\ 
\includegraphics[width=0.43\textwidth]{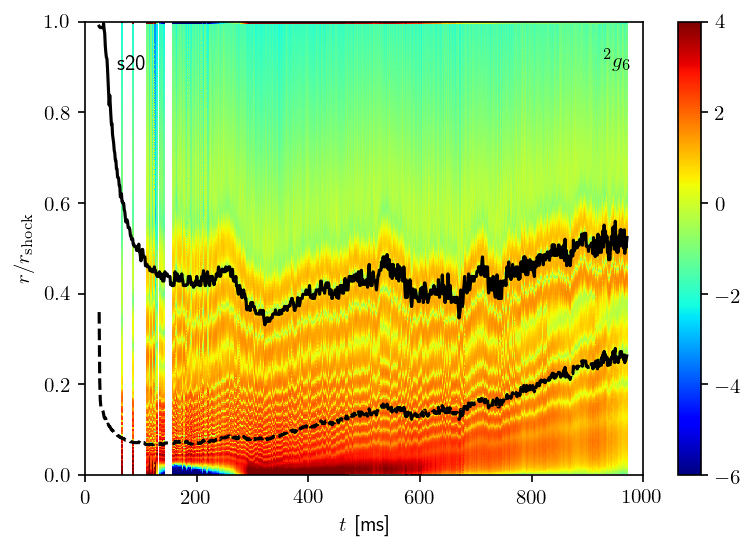}
\includegraphics[width=0.43\textwidth]{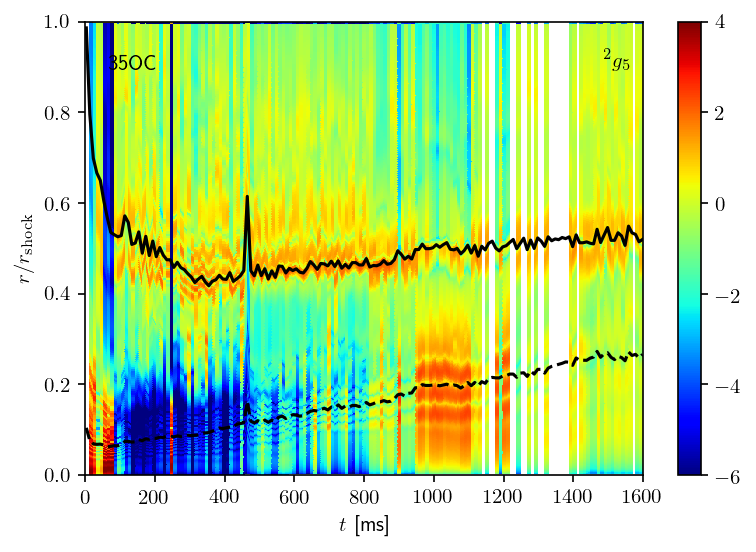}
\caption{Same as Fig.~\ref{fig:morpho_p}, but for a selection of modes classified as surface g-modes. Note that in the model 35OC
some modes have been misclassified by our algorithm and appear as sharp transitions.}
\label{fig:morpho_gsurf}
\end{figure*}

\begin{figure*}
\includegraphics[width=0.45\textwidth]{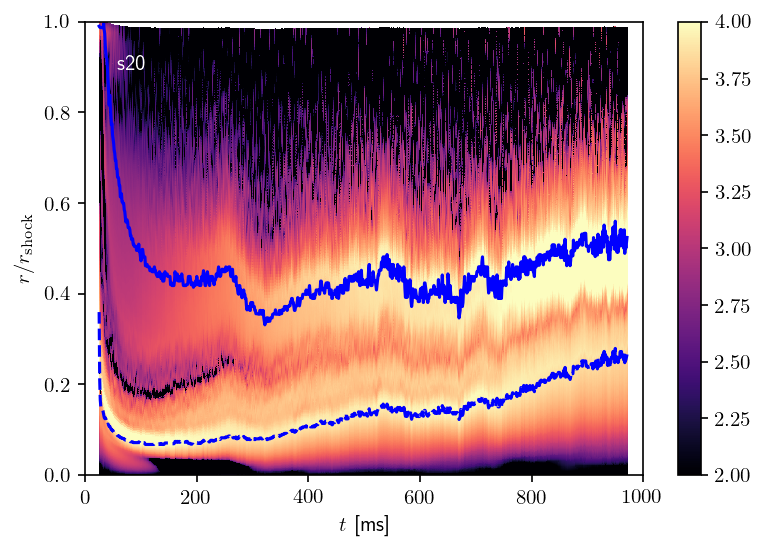}
\includegraphics[width=0.45\textwidth]{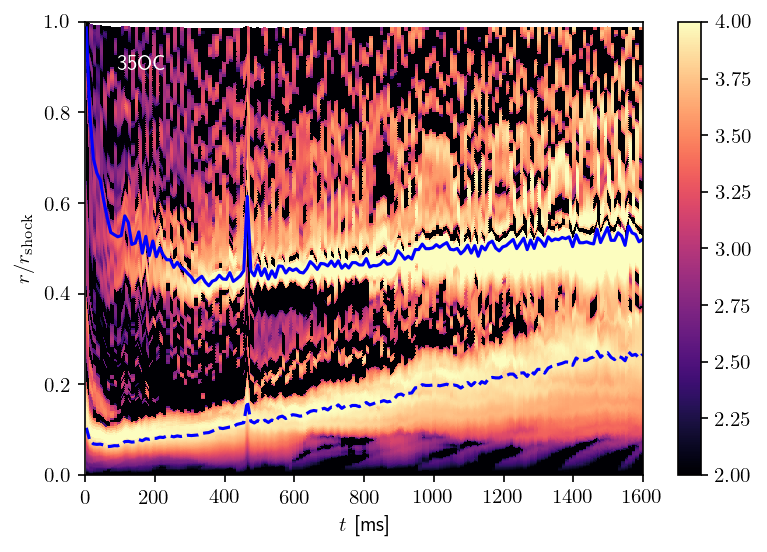}
\caption{Logarithm  of the Brunt-V\"ais\"al\"a frequency, $\mathcal{N}$, for positive values of $\mathcal{N}^2$, i.e. regions stable against convection. Convectively unstable regions ($\mathcal{N}^2<0$)  are ploted in black. The blue black line indicates the position of the PNS surface and the dashed blue line the surface of the  inner core. The radius is normalised to the shock location.}
\label{fig:BV}
\end{figure*}
 
\subsection{Eigenmode morphology}
 
Figures~\ref{fig:morpho_p} to \ref{fig:morpho_gsurf} show the time evolution of the Newtonian eigenmode energy density, defined 
by Eq.~(\ref{eq:energyden2}), for a selection of modes classified in three groups 
according to their shape:
\begin{itemize}
\item {\it f-mode and p-modes}: Fig.~\ref{fig:morpho_p} shows a selection of modes ($^2f$, $^2p_1$, $^2p_2$, $^2p_3$) with a relatively large
amount of energy density outside the PNS. These modes are basically trapped sound waves in the region between the PNS surface
and the shock. However, due to the complex hybridisation process mentioned above, they also couple with regions deep in the PNS interior, specially
at frequencies where mode crossings appear. For higher values of $n$ the number of nodes (here appearing as deeps in $\mathcal E$) increases
as expected for higher order overtones of the f-mode. All these modes have in common that their frequencies increase in time (except for some 
periods in the s20 model) and that the frequencies are integer multiples of the f-mode frequency.
\item {\it Core g-modes}:  Fig.~\ref{fig:morpho_gcore}  collects modes with the largest amplitude inside the PNS ($^2g_1$, $^2g_2$ and $^2g_3$).
While lower $n$ modes are more extended over the whole PNS, higher order modes become more concentrated in the inner core region.
These modes correspond to g-modes associated with the stable region in the innermost part of the PNS  (see Fig.~\ref{fig:BV}).
The frequency evolution of these modes is similar for both models. While the $^2g_1$ mode shows an almost monotonic increase in frequency,
higher order overtones ($^2g_2$ and $^2g_3$) decrease their frequency with time. 
\item {\it Surface g-modes}: Fig.~\ref{fig:morpho_gsurf} collects modes with the largest amplitude at the surface of the PNS (e.g. $^2 g_4$).
These modes result of the excitation of the buoyantly stable layer ($\mathcal{N}^2>0$) at the 
surface of the PNS  (see Fig.~\ref{fig:BV}). These modes show an almost monotonically increase of their frequency during their evolution, 
although they are confined to low frequencies ($<250$~Hz).
\end{itemize}

This interpretation is consistent with previous analysis (albeit more simplified) of the PNS structure \citep{Cerda-Duran:2013,Andresen:2017}, which 
showed that the features observed in the GW spectrograms could be related to g-modes in two different regions (surface and inner core). 

\section{Comparison with simulations}
\label{sec:simulations}

\subsection{GW spectrograms}
\label{sec:spec}

\begin{figure*}
\flushleft
\includegraphics[width=0.425\textwidth]{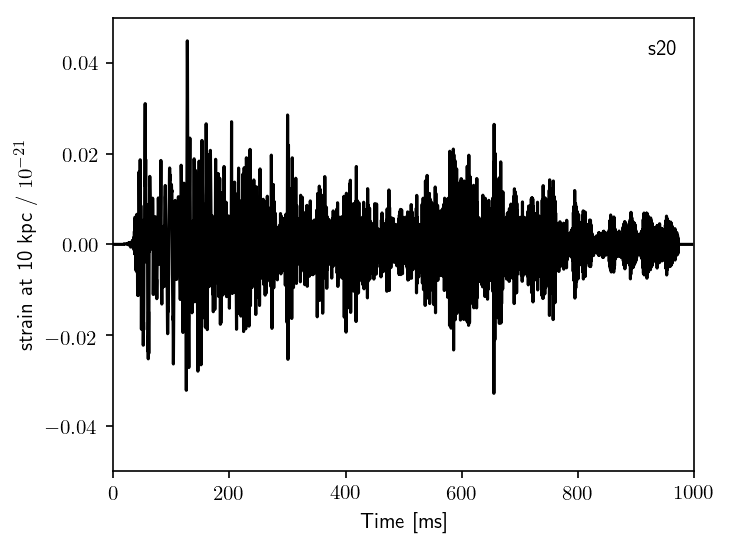} \hspace{0.06\textwidth}
\includegraphics[width=0.425\textwidth]{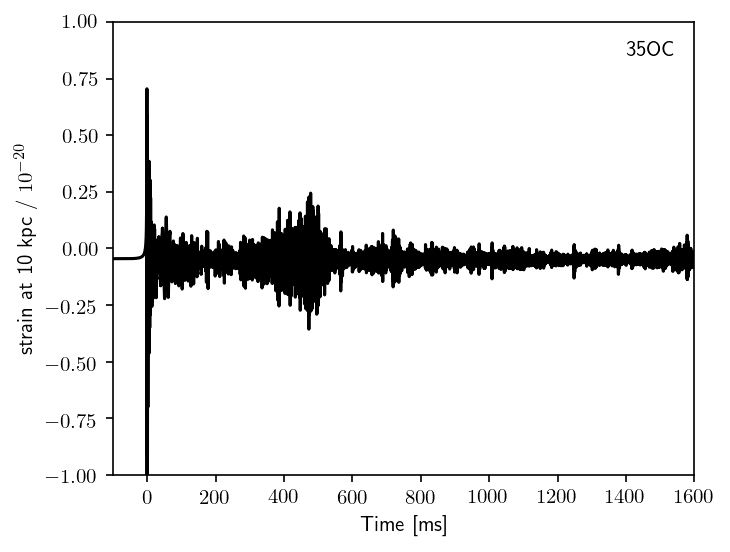}\\
\includegraphics[width=0.49\textwidth]{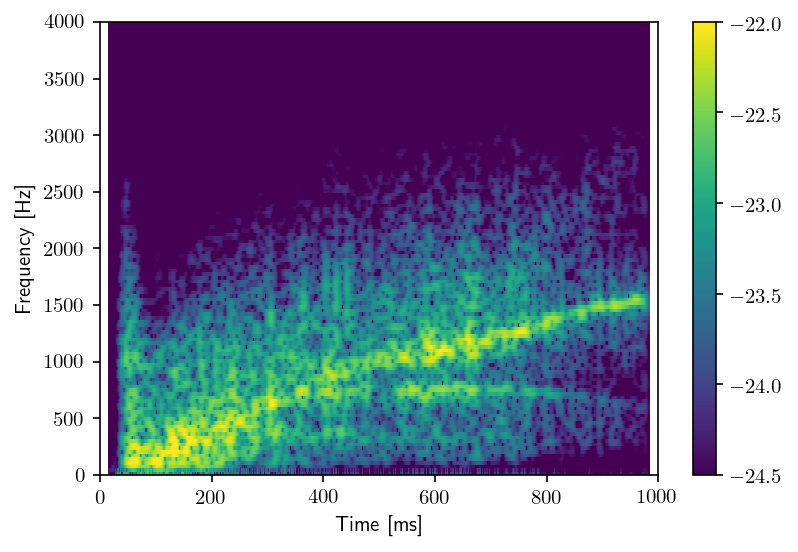}
\includegraphics[width=0.49\textwidth]{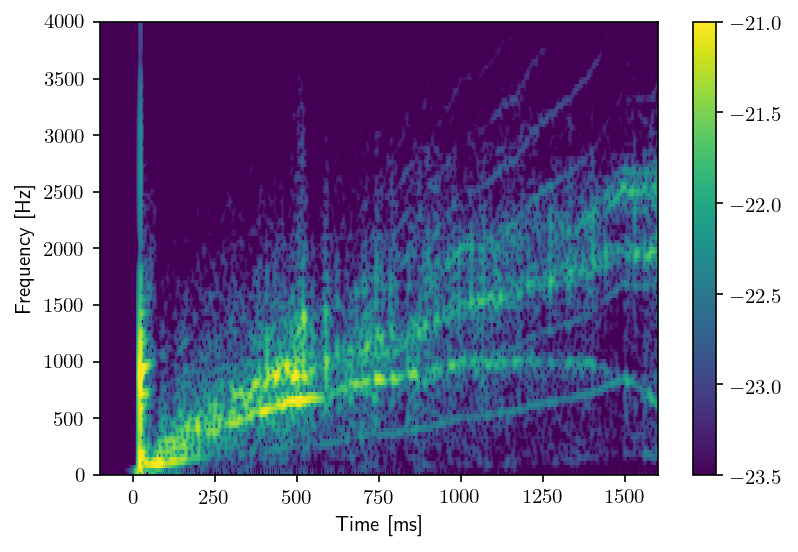}\\
\includegraphics[width=0.49\textwidth]{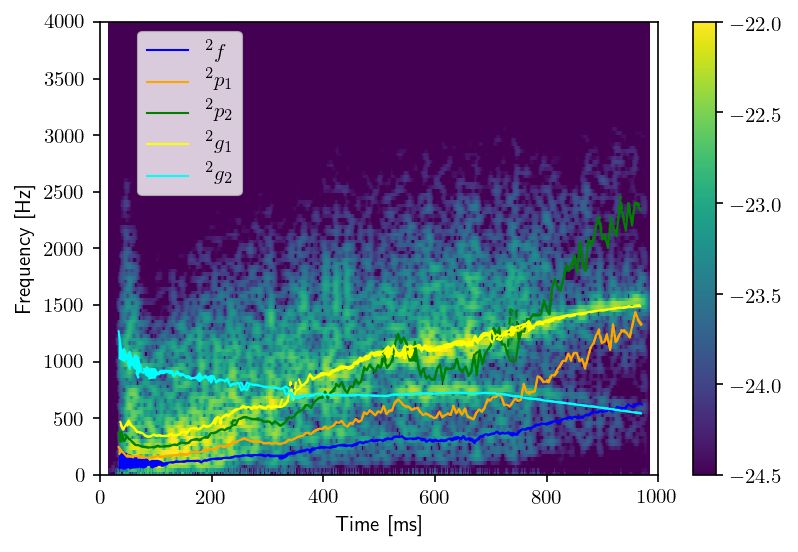}
\includegraphics[width=0.49\textwidth]{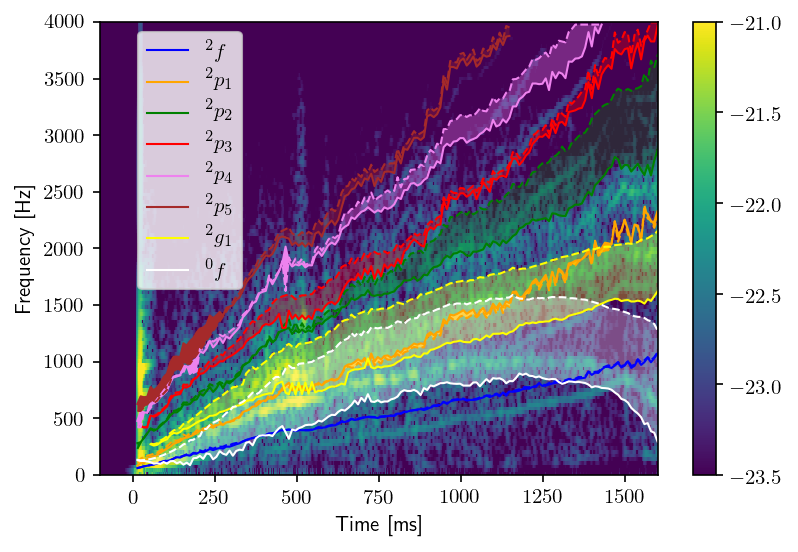}
\caption{This figure shows the
GW signal (upper panels), the corresponding 
spectrograms (middle panels), and the spectrograms with a selection of modes over-plotted (lower panels) for models s20 (left) and 35OC (right). Solid lines and dashed lines are used to indicate that the calculations were made using $\mathcal G_P$ and $\mathcal G_\alpha$, respectively. Note that for model s20 those two lines overlap.}
\label{fig:GWcomp}
\end{figure*}

In this section we compare the eigenmode frequencies obtained applying the linear analysis to the 
spherical background with the GW frequencies computed in the numerical simulations, result of the multidimensional
dynamics of the model. Fig.~\ref{fig:GWcomp} shows the GW signal (upper panels), the corresponding 
spectrograms (middle panels) and the spectrograms with a selection of modes over-plotted (lower panels).
As mentioned before, there is some uncertainty in the definition  of $\mathcal G$. To gauge the difference 
between using $\mathcal G_P$ and $\mathcal G_\alpha$, we plot in all cases the 
modes computed in both ways. The range between both options gives a measure of the error introduced
by the definition of $\mathcal G$.

For the non-rotating s20 model (left panels), both calculations lay on top of each other. This is a strong indication that the assumption of hydrostatic equilibrium inside the shock is indeed a very good approximation. However, in the rapidly-rotating model 35OC (right panels), there is a significant difference between both definitions of $\mathcal G$ in many of the models (solid lines correspond to $\mathcal G_P$ and dashed lines to $\mathcal G_\alpha$). The difference may be due to the fact that we are neglecting rotation in our linear analysis. In fact, centrifugal forces play an important role in the equilibrium for this model and, thus, this result is not a surprise. What is reassuring (and somewhat striking) is that, even using a non-rotating analysis, the features observed in the spectrogram lay between our two possibilities for the eigenmode computation. 

In general, for both models (see lower panels of Fig.~\ref{fig:GWcomp} there is a clear agreement of a selection of modes with the features in the spectrograms, which is a clear indication that these modes are responsible for the computed GW emission. In model s20 (bottom left panel) the agreement for modes $^2g_1$ and $^2g_2$ is remarkable with some hints of the presence of the $^2f$ mode. Although p-modes are not clearly visible, some excess power above the $^2g_1$ mode 
may be an indication of their presence. In model 35OC (bottom right panel), all main features can be explained with a few modes. Most of them match well the spectrogram within the error produced by the definition of $\mathcal G$. The only exception is the $^2f$ mode, whose evolution runs in parallel to the spectrogram feature but with higher frequency. The main features can be explained by the $^2g_1$ mode and the $^2p_1$ mode. The f-mode and all p-modes up to order 5 are also clearly visible, albeit with lower amplitudes. We note in particular that our computation of the $l=0$ mode is able to reproduce the characteristic feature of this mode close to black hole formation, namely that its frequency goes to zero at the onset of instability~\citep{Cerda-Duran:2013}, as predicted 
by~\cite{Chandrasekhar:1964}.

In addition to estimating the effect of the definition of $\mathcal G$ in our mode comparison, we also test its effect in the expression for the Brunt-V\"ais\"al\"a
frequency. In this work we first perform an angular average of the simulation data and then we compute the Brunt-V\"ais\"al\"a frequency as $\mathcal N^2 = \mathcal{G} \mathcal{B}$, being  $\mathcal{G}$ and $\mathcal{B}$ the radial component of  the vectors $\mathcal{G}_i$ and $\mathcal{B}_i$. Alternatively one can compute  $\mathcal N^2 = \mathcal{G}_i \mathcal{B}^i$, on the 2D grid of the simulation and then perform the angular average to obtain 1D profiles of $\mathcal N^2$. For the fast rotating case, the second procedure takes into account the non-radial components of  $\mathcal{G}_i$ and $\mathcal{B}_i$, which are otherwise neglected in the first procedure. We have computed the eigenmodes using both definitions and the differences in the computed eigenfrequencies do not differ by more than $1\%$.

\begin{figure*}
\flushleft
\includegraphics[width=0.48\textwidth]{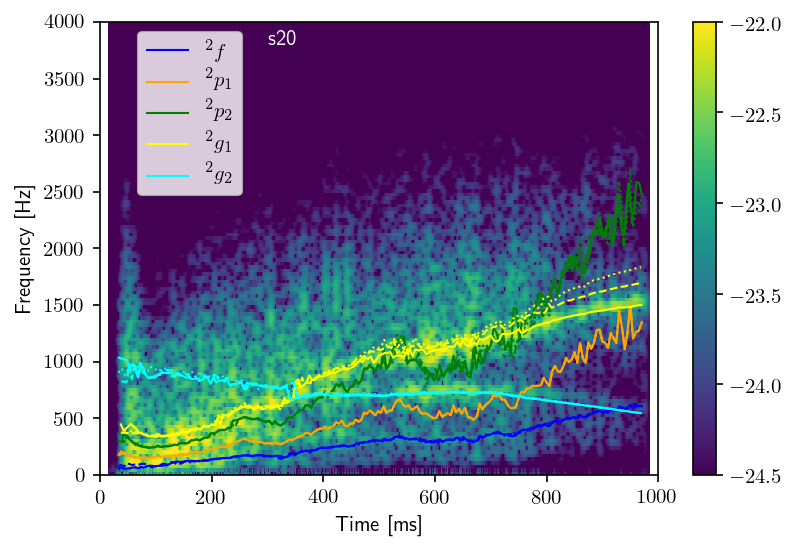}
\includegraphics[width=0.48\textwidth]{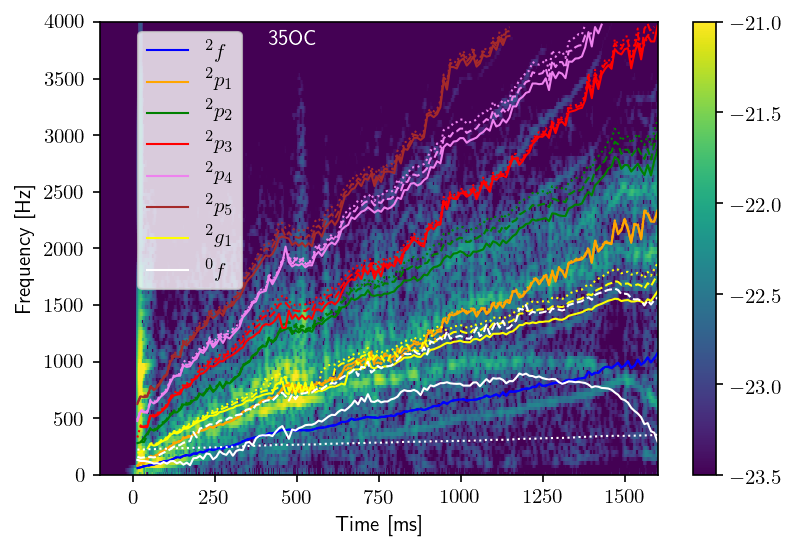}
\caption{Spectrograms with a selection of modes over-plotted using spacetime perturbations (solid lines), 
only metric perturbations of $\delta \hat Q$ (dashed lines), and no metric perturbations (Cowling, dotted lines), 
for models s20 (left) and 35OC (right).}
\label{fig:gravity}
\end{figure*}

Regarding gravity, this work improves over previous work that considered no metric perturbations \citep[Cowling, ][]{Torres-Forne:2018}
and only perturbations of the lapse function \citep{Morozova:2018}. To compare with different approaches for the metric perturbations we have
performed our analysis in Cowling and considering only perturbations of $\delta \hat Q$, which corresponds to including only perturbations of the lapse function.
Fig.~\ref{fig:gravity} compares the three approaches. In general, the differences among them increase at later times. This is because the
PNS becomes gradually more compact and GR effects become more relevant. Note in particular the $^2g_1$ mode for the s20 model
(yellow lines in the left panel), in which only the approach followed in the present work is able to reproduce the features observed in the spectrogram. For the rotating model 35OC, this is more difficult to discuss, because the differences with respect to the spectrogram are also strongly influenced by rotation.
However, it is worth highlighting that our approach is the only capable of reproducing the turning down of the $^0f$ mode at the onset of black hole formation, while  the other two approaches give qualitatively different behaviors. Therefore, we conclude that our approach is necessary whenever high accuracy in the
eigenmode calculations is needed, specially close to black hole formation. Moreover, the high accuracy obtained in model s20 provides convincing evidence that
perturbations of the shift vector, neglected in our analysis, are not important, at least for slowly-rotating systems.

\subsection{GW efficiency and mode energy}

\begin{figure}
\flushleft
\includegraphics[width=0.48\textwidth]{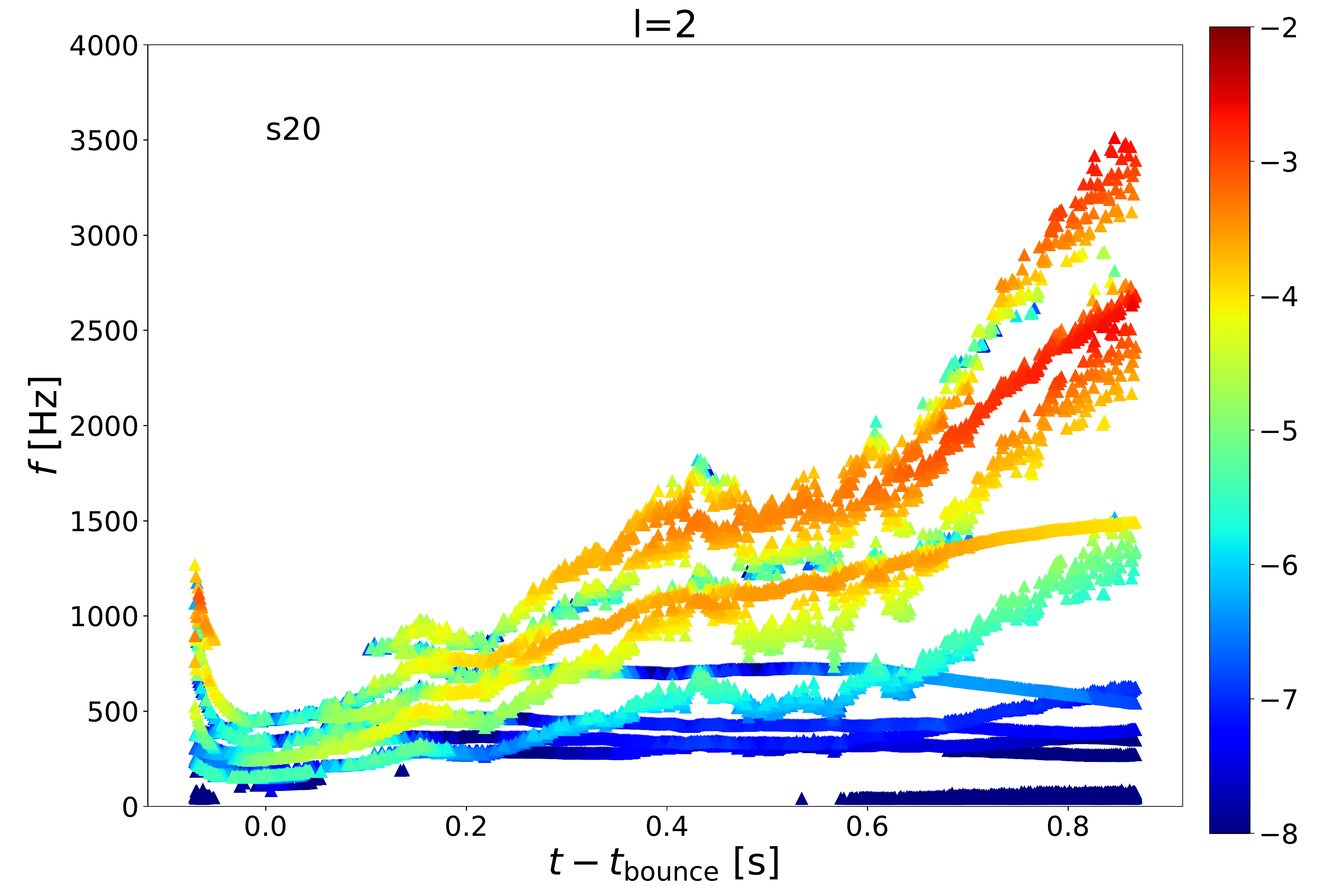}
\includegraphics[width=0.48\textwidth]{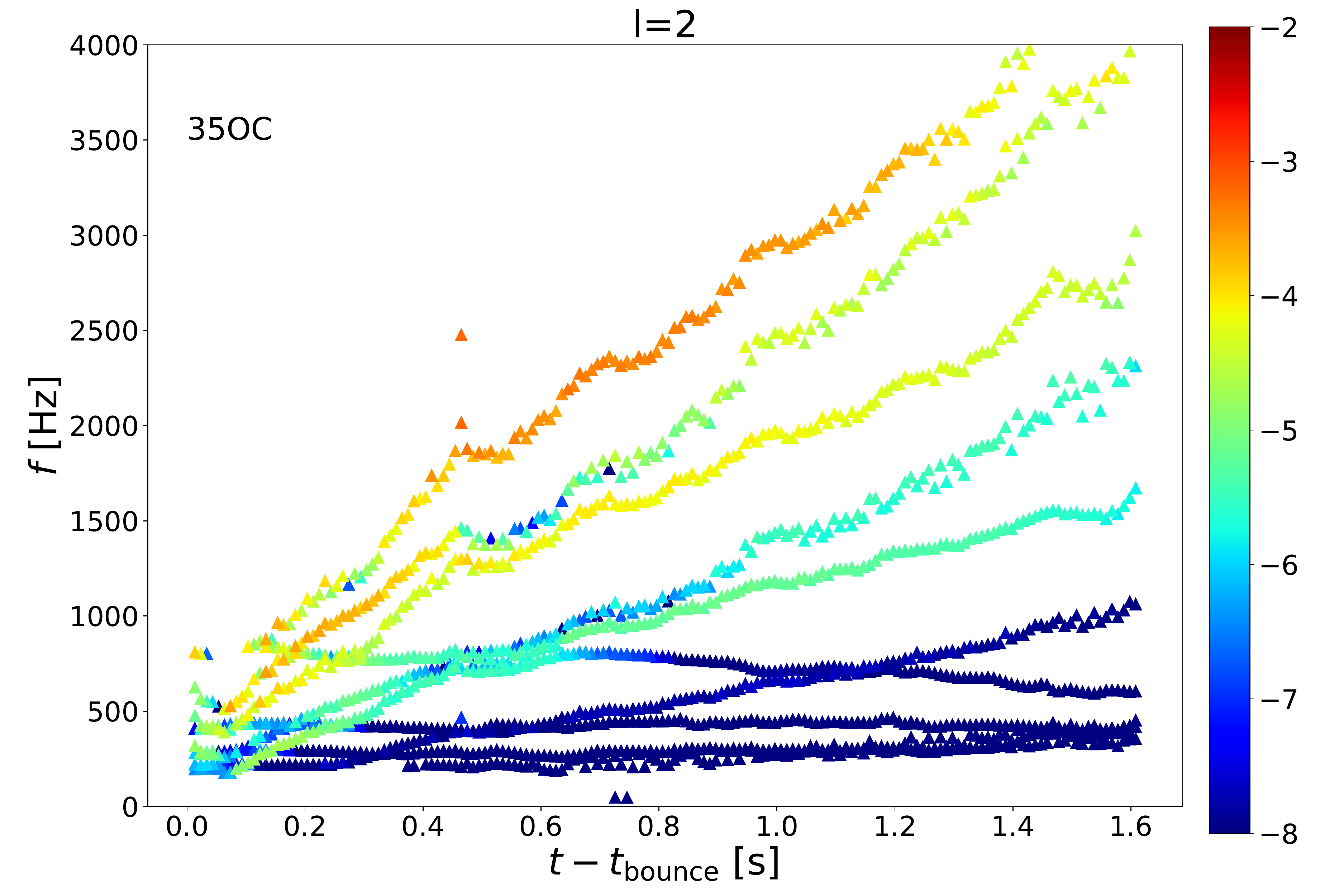}
\includegraphics[width=0.48\textwidth]{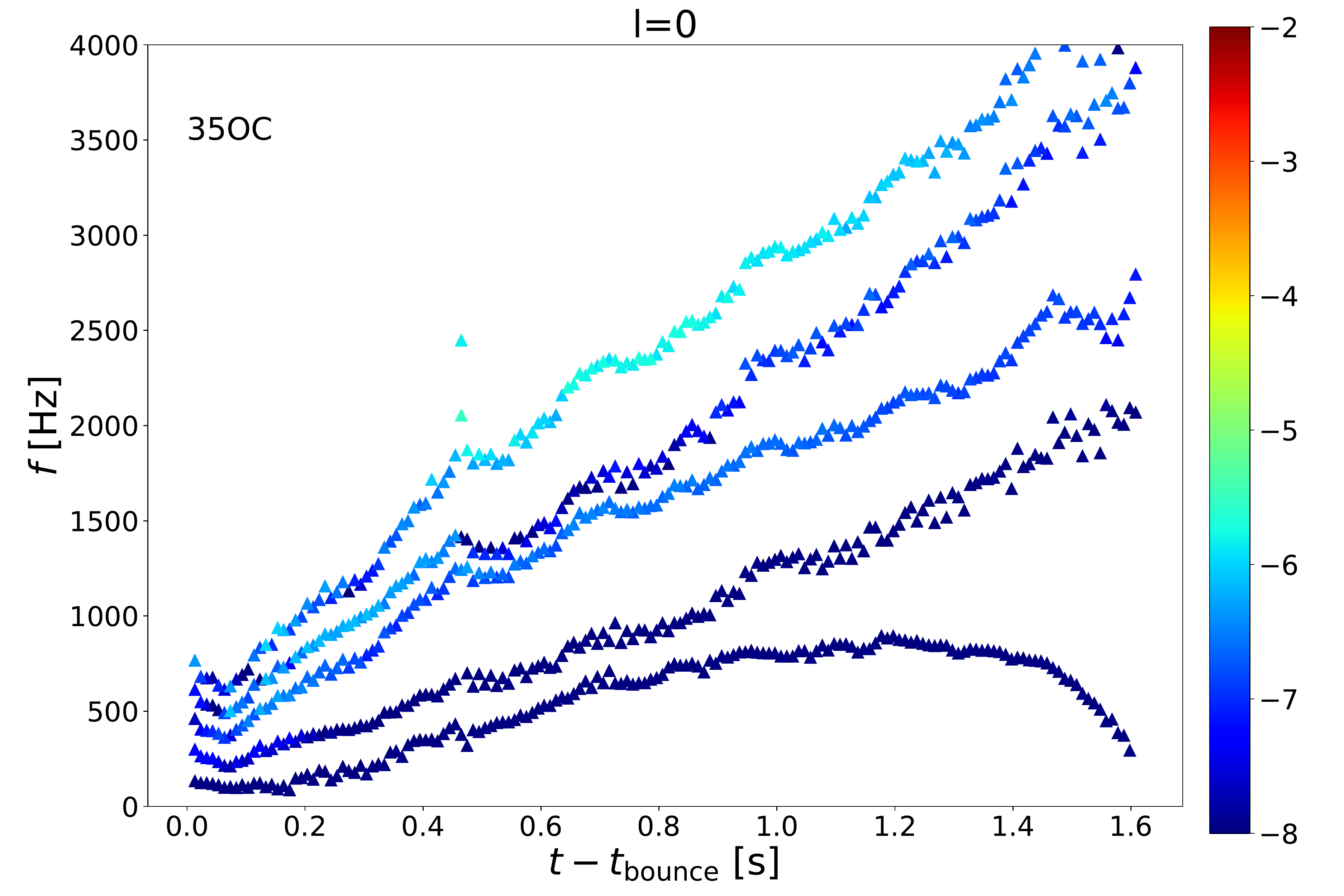}
\caption{GW efficiency for the $l=2$ modes in model s20 (upper panel) and for the $l=2$ and $l=0$ modes in model 35OC
(middle panel and lower panel, respectively).}
\label{fig:efficiency}
\end{figure}

To understand which modes are more efficient GW emitters, we compute the GW efficiency (see Eq.~(\ref{eq:effi})) for $l=2$ and $l=0$ modes. This calculation provides an estimation of the fraction of the energy stored in an eigenmode that is emitted per oscillation cycle. Model s20 is non rotating, so the $l=0$ mode will not contribute to the GW signal. Model 35OC corresponds to a rapidly rotating progenitor, which results in a PNS with a central period of about $1.5$~ms at bounce. The PNS is considerably deformed, with a polar-to-equatorial radius ratio of $~0.64$ after bounce and the GW amplitude is expected to be significantly affected by rotation (see Section~\ref{sec:deformed}). We use Eq.~(\ref{eq:dlm}) to estimate the quadrupolar moment needed for the computation of the GW efficiency. In our numerical simulation the ratio $r_{\rm e}/r_{\rm p}$ ranges from $\sim 1.25$ (in the PNS core) to $\sim 1.7$ (at the PNS surface), which correspond to elllipticities in the range $e \sim 0.6 - 0.8$. The pre-factor to the integral of Eq.~(\ref{eq:dlm}) is in the interval $-0.06$-$-0.14$ for our ellipticity range. This means that, for similar eigenfunction structure and mode energy, the $l=0$ modes are expected to emit GWs with an amplitude of about $10\%$ the amplitude of the $l=2$ modes.

Fig.~\ref{fig:efficiency} shows the GW efficiency for both models. Similarly to the results of \cite{Torres-Forne:2018}, the efficiency 
grows with the frequency, with the p-modes being in general more efficient than the g-modes. This happens because the GW efficiency
approximately scales with $\sigma^3$. The implication is that to see low-frequency modes in the spectrogram one needs considerably
large energy in those modes. In both models the most visible features correspond to the lowest order modes, while high order p-modes
are only observed in the 35OC model, likely related to the SASI activity (see the related discussion in Section~\ref{sec:summary}). In neither model
high order g-modes are observed, the highest order mode observed being the $^2g_2$ of the s20 model.

\begin{figure}
\flushleft
\includegraphics[width=0.48\textwidth]{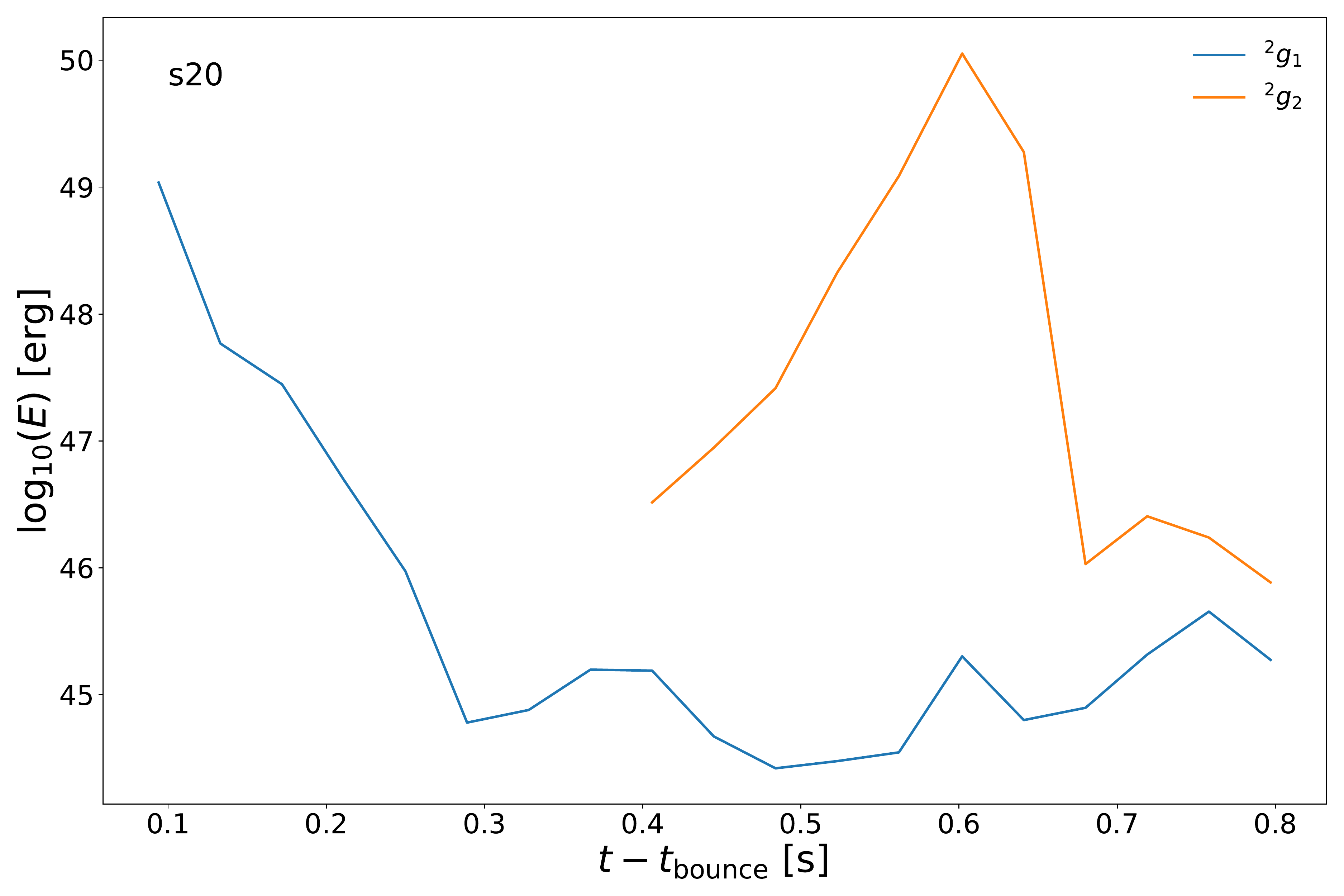} 
\includegraphics[width=0.48\textwidth]{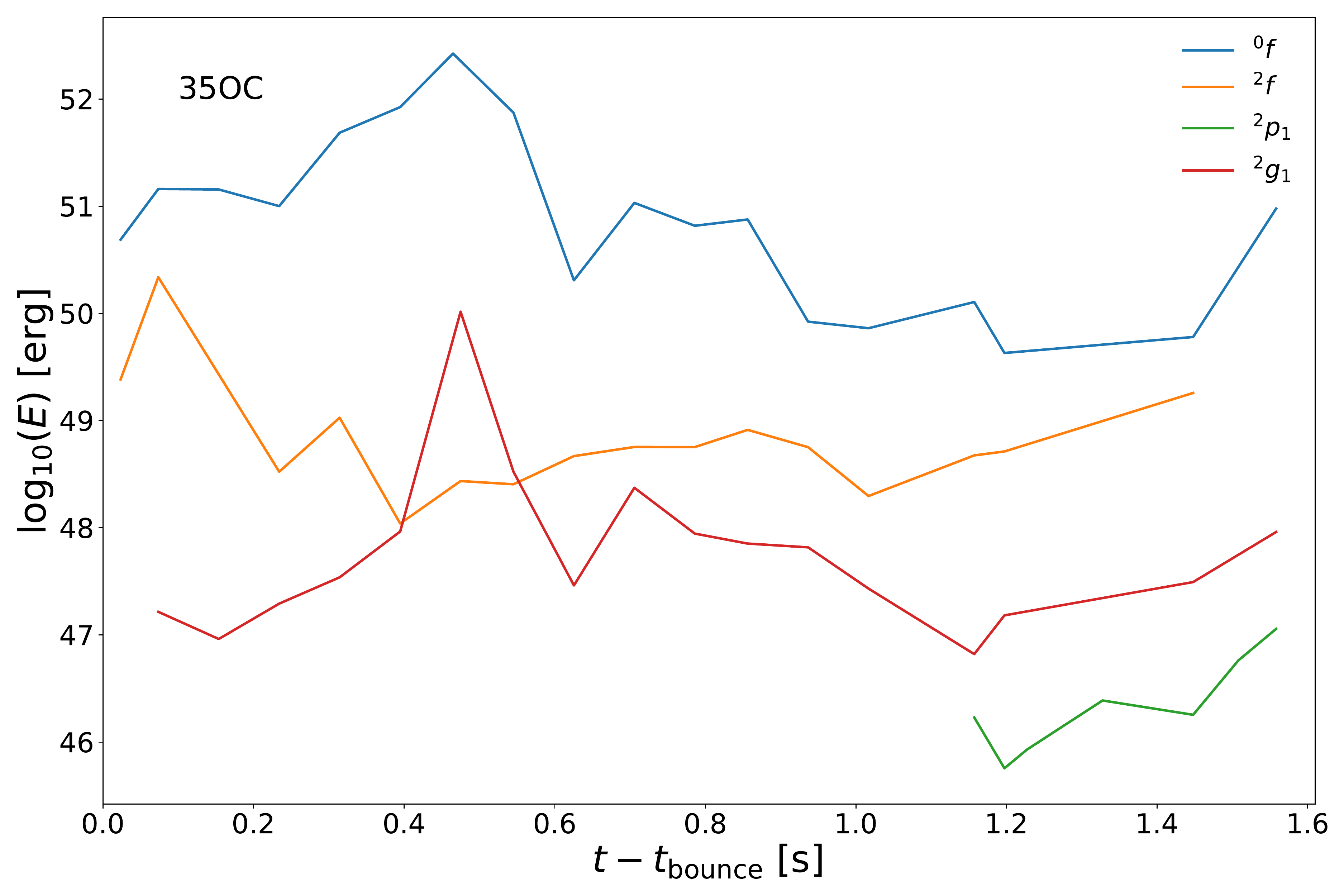}
\caption{Energy of the principal emission modes in logarithmic scale for $l=2$ modes in model s20 (upper panel) and for $l=2$ and $l=0$ modes in model 35OC
(lower panel).}
\label{fig:eps_hplus}
\end{figure}

Given that we know the complete eigenmode structure, we can use the spectrogram to estimate the energy stored in each mode
and try to validate the previous claims. To this end, we extract the amplitude of the GW emission at some intervals along the main features of the spectrogram. Then, we rescale the amplitude of the eigenmodes so that their GW emission amplitude computed with Eq.~(\ref{eq:h+}) matches 
the value obtained from the spectrogram. The results of this calculation for model s20 are shown in the upper panel of Fig.~\ref{fig:eps_hplus}. The modes which correspond to the main features on the 
spectrogram (Fig.~\ref{fig:GWcomp} middle left panel) are the $^2g_1$ and the $^2g_2$. The highest power in GW corresponds to the $^2g_1$ mode, while the other mode presents a slightly less amplitude. However, the energy stored in each mode (Fig.~\ref{fig:eps_hplus}) exhibits the opposite behaviour. This is due to the GW efficiency 
(Fig.~\ref{fig:efficiency}  upper panel). As the efficiency of the $^2g_2$ mode is much lower than that of the $^2g_1$ mode, the corresponding energy should be larger to radiate a similar energy in GW. In the case of model 35OC (Fig.~\ref{fig:eps_hplus} lower panel), we have analyzed the four modes which show the most clear trace in the spectrogram (Fig.~\ref{fig:GWcomp} middle right panel). The energies are ordered as expected. The two fundamental modes, $^0f$ and $^2f$, have the largest energies because their GW amplitude are large (their traces on the spectrogram are clearly visible) but their efficiencies are low. 

\section{Summary and discussion}
\label{sec:summary}

In this work we have developed a numerical procedure to solve the eigenvalue problem of hydrodynamic and metric 
perturbations of a spherically symmetric self-gravitating system in general relativity. We have applied this method to
compute the oscillation frequencies of the PNS-shock system formed during a core-collapse supernova explosion. Those frequencies have been compared
to the time-frequency patterns observed in the GW templates from two CCSN numerical simulations. This work is an extension 
of our previous investigation~\citep{Torres-Forne:2018} and brings forth significant improvements with respect to previous results. 
The numerical routines developed and used in this work to solve the eigenvalue problem are available at the {\tt GREAT} library
(General Relativistic Eigenmode Analysis Tool) that we have released as open source 
({\tt https://www.uv.es/cerdupa/codes/GREAT/}).

The main results of this work can be summarized as follows:

\begin{itemize}

\item We have incorporated perturbations of the lapse function and of the conformal factor, improving the approach of 
\cite{Torres-Forne:2018}, based on the Cowling approximation, and of \cite{Morozova:2018}, which considered only
perturbations of the lapse function. Our results show that it is necessary to consider both kinds of perturbations to
 accurately reproduce the $l=2$ modes in the GW spectrograms. Regarding $l=0$ modes, our approach is the only one 
able to trace the qualitative time-frequency behaviour of the $^0f$ mode, for which the frequency decreases towards zero at the onset
of black hole formation. This mode is of particular importance because  it indicates the formation of a black hole and could be observed for rapidly rotating cores. Our approach has not accounted for perturbations of the shift vector. However, given the accuracy of our results, incorporating those perturbations does not seem necessary, at least for non-rotating models.

\item For our nonrotating model s20, we have found an excellent agreement between the features observed in the spectrogram and
the computed eigenmodes. For rapidly rotating cores, the assumed spherically-symmetric background
is not valid and the results are only qualitatively similar to the GW signal, due to the effect of centrifugal forces, not
considered in our analysis. 

\item In both CCSN models, s20 and 35OC, the $^2g_1$ mode has been identified as the main contributor to the GW signal. The $^2f$ mode is also visible
in the two models, along with a few overtones. 
Furthermore, we have estimated the eigenmode energy according to the amplitude of the GW signal for each mode. This analysis 
confirms that most of the energy is stored in the lowest order eigenmodes.

\item We have developed a formalism to estimate the contribution of quasi-radial modes ($l=0$)
to the quadrupolar component of the GW signal in the case of deformed backgrounds. This is of particular importance
for rapidly rotating cores in which the PNS is deformed by the effect of centrifugal forces.

\item Our improved analysis has been possible thanks to a newly developed matching classification scheme, that allows to classify
the eigenmodes in p-, g- and f-modes, in a more accurate way as the preceding methods used in~\citet{Torres-Forne:2018}. 

\end{itemize}

Despite our investigation has been limited to only two numerical models, some conclusions  about the mechanism producing GWs during the collapse of massive stars can be extracted from this work, in particular when comparing with previous results in the literature. The presence of a clear pattern of rising frequencies in the GW spectrograms of the post-bounce evolution of PNSs has been observed in a number of works \citep{Murphy:2009,Mueller:2013,Cerda-Duran:2013,Yakunin:2015,Kuroda:2016,Andresen:2017}. In all these works, this raising pattern was attributed to g-modes of the PNS. Our work shows that this main feature is probably the $^2g_1$ mode, i.e.~the lowest order g-mode, associated with the buoyantly stable region in the innermost part of the PNS, and not to its surface. Although surface g-modes are possible according to our analysis, we do not see their correspondence in the GW spectrograms of the two models computed. This is very likely due to the low GW efficiency of these modes, which displace a much smaller amount of matter when compared with the core g-modes. 

Some works in the literature \citep{Cerda-Duran:2013,Kuroda:2016,Andresen:2017} have related the presence of a low frequency ($\sim 100$~Hz)
component in the GW signal to the characteristic frequency of the SASI. In this work, however, we have identified the mode observed
in \cite{Cerda-Duran:2013} as the fundamental $^2f$ mode, and we suspect a similar result will hold for the results of \cite{Kuroda:2016} and \cite{Andresen:2017}. The fundamental mode (and also higher order p-modes) seems to be excited in cases with strong SASI activity. \cite{Cerda-Duran:2013} showed that these features
observed in the GW spectrogram match perfectly with the features observed in the spectrogram of the time evolution of the shock, i.e.
the shock oscillates with frequencies matching the p-modes. It is not completely clear why this is the case \citep[see discussion in][]{Cerda-Duran:2013}.
In principle, in the presence of the SASI the shock oscillates with frequencies corresponding to the unstable modes of the advective-acoustic cycle
coupling the shock and the PNS surface \citep{Foglizzo:2002, Foglizzo:2007}, which should not coincide with the frequencies of the purely acoustic
cycle (p-modes in our case). We are not sure if the matching of the frequencies observed in our analysis is generic or if it is a particular feature of the 35OC model
(e.g.~a resonance). The analysis presented in this work can be applied to other multidimensional simulations presenting signatures of SASI to
try to better understand the relationship between p-modes and SASI.

Our clear identification of the $^2g_1$ and $^2f$ modes is of great importance to devise strategies to infer PNS properties from a possible GW observation
of a nearby SN (or BH formation) event. It follows from our work that most of the future efforts must focus in these two particular modes,
in order to learn how they depend on the properties of the PNS. This will also allow to develop data analysis methods aimed at detecting this kind
of GW signals buried in detector noise. 

Our analysis may also simplify the GW templates from CCSNe, since most of the information (the time-frequency behaviour) encodes the general evolution of the PNS, and not  the particular non-linear perturbations,
which are to a large degree stochastic. Some of the information cannot be extracted from our analysis, most notably the energy stored in each of the eigenmodes, which does not allow us to provide complete templates of CCSNe. For this we still will have to rely on sophisticated core-collapse simulations. One of the main questions that has to be addressed is what is the
evolution of the energy content of each eigenmode. The results presented here are only for two 2D simulations, but the energy 
distribution may change significantly in 3D simulations or in simulations accounting for more sophisticated microphysics or neutrino transport.

Finally, we note that the results presented in this work strongly rely in our new eigenmode classification algorithm. Previous results by \cite{Cerda-Duran:2013}
suffered from serious misclassification issues. In particular, the so-called SASI mode was not classified as the fundamental mode, which arose confusion to explain why it was an integer division of the higher order modes. This is however explained naturally with its identification as an f-mode, possible with the new matching classification scheme. This method is not completely automatic and requires some degree of manual intervention. We hope to improve our classification method in the future to allow for a fully automatic and robust eigenmode classification procedure.

\section*{Acknowledgements}
Work supported by the Spanish MINECO (grant AYA2015-66899-C2-1-P), by the Generalitat Valenciana (PROMETEOII-2014-069) and by the European Gravitational Observatory (EGO-DIR-51-2017). P.C~ acknowledges the support from the Ramon y Cajal program of the Spanish MINECO (RYC-2015-19074).
A.P.~acknowledges support from the European Union under the Marie Sklodowska Curie Actions Individual Fellowship, grant agreement no 656370. J.A.F.~acknowledges support from  the  European  Union's  Horizon  2020  RISE programme H2020-MSCA-RISE-2017 Grant No.~FunFiCO-777740.



\bibliographystyle{mnras}
\bibliography{./references}



\appendix 

\section{Coefficients of the perturbative equations}
\label{sec:appendix:coeff}

The system of equations described in Sections~\ref{sec:l2} and \ref{sec:l0} (for $l\ne0$ and $l=0$, respectively) can be cast in a matrix form as
\begin{equation}
\partial_r \bm u = \bm A \bm u.
\end{equation}

For $l\ne0$, $\bm u \equiv (\eta_r,\eta_\perp, \fq,\delta\hat Q,\fpsi,\delta\hat\psi)^{T}$ and the non-zero elements of $\bm A$ are given by
\begin{align}
A_{11} & =  - \frac{2}{r} - \frac{ \G}{c_s^2} - 6 \frac{ \partial_r \psi }{\psi}\,, \nonumber \\
A_{12} &= \frac{\psi^4 }{\alpha^2 c^2_s} \left( \mathcal{L}^2 - \sigma^2\right)\,, \nonumber \\
A_{14} & = \frac{1}{c_s^2 Q}\,, \nonumber \\
A_{16} & = - \left ( 6 + \frac{1}{c_s^2} \right )\frac{1}{\psi}\,, \nonumber 
\end{align}
\begin{align}
A_{21} & = \left( 1 - \frac{ \mathcal{N}^2 }{\sigma^2} \right)\,, \nonumber \\
A_{22} & = -\partial_r \ln q + \G \left( 1 + \frac{1}{c^2_s} \right)\,, \nonumber \\
A_{24} & =   \frac{\alpha}{\psi^5\sigma^2} \left [ \partial_r (\ln \rho h) - \left ( 1 + \frac{1}{c_s^2} \right )\G \right ]\,, \nonumber \\
A_{26} & =   -\frac{\alpha^2}{\psi^5\sigma^2} \left [ -\partial_r (\ln \rho h) + \left ( 1 + \frac{1}{c_s^2} \right )\G \right ] = -\alpha A_{24}\,,\nonumber 
\end{align}
\begin{align}
A_{31} & = -2\pi \rho  h \alpha \psi^5 \mathcal{B}\,, \nonumber \\
A_{32} & = 2\pi \rho  h \alpha \psi^5 \left ( 6 + \frac{1}{c_s^2}\right ) \frac{\psi^4 \sigma^2}{\alpha^2}\,, \nonumber \\
A_{33} & = -\frac{2}{r}\,, \nonumber \\
A_{34} &= \frac{l(l+1)}{r^2} - 2\pi\psi^4\left (  5 e + \frac{\rho h}{c_s^2}\right )\,, \nonumber \\ 
A_{36} & = 2 \pi \psi^4 \alpha \left [ 10 e + 30 P + \frac{\rho h}{c_s^2} \right ]\,, \nonumber 
\end{align}
\begin{align}
A_{43} & = 1\,, \nonumber 
\end{align}
\begin{align}
A_{51} & = 2 \pi \rho h \psi^5  \mathcal{B}\,, \nonumber\\
A_{52} & = -2 \pi \rho h \psi^5  \frac{\psi^4 \sigma^2}{\alpha^2 c_s^2}\,, \nonumber \\
A_{54} & = 2 \pi \psi^4 \frac{\rho h}{\alpha c_s^2}\,, \nonumber \\
A_{55} & = -\frac{2}{r}\,, \nonumber \\
A_{56} & = \frac{l(l+1)}{r^2} - 2 \pi \psi^4 \left ( 5 e + \frac{\rho h}{c_s^2} \right )\,, \nonumber \\
A_{65} & = 1\,. \nonumber
\end{align}
For $l=0$, $\bm u \equiv (\eta_r,\delta\hat P, \fq,\delta\hat Q,\fpsi,\delta\hat\psi)^{T}$ and the non-zero elements of $\bm A$ are
\begin{align}
A_{11} & =  - \frac{2}{r} - \frac{ \G}{c_s^2} - 6 \frac{ \partial_r \psi }{\psi}\,, \nonumber \\
A_{12} &= - \frac{1}{\rho h c_s^2}\,, \nonumber \\
A_{16} & = -\frac{6}{\psi}\,, \nonumber 
\end{align}
\begin{align}
A_{21} & =-q (\mathcal{N}^2 - \sigma^2)\,, \nonumber \\
A_{22} & = \G \left( 1 + \frac{1}{c^2_s} \right)\,, \nonumber \\
A_{23} & = -\rho h \alpha^{-1} \psi^{-1}\,, \nonumber \\
A_{24} & = \rho h \alpha^{-1} \psi^{-1} (\partial_r \ln \psi -\G)\,,   \nonumber \\
A_{25} & =   \rho h \psi^{-1}\,, \nonumber \\
A_{26} & =  -\rho h \psi^{-1} (\partial_r \ln \psi)\,, \nonumber 
\end{align}
\begin{align}
A_{31} & = -2\pi \rho  h \alpha \psi^5 \mathcal{B}\,, \nonumber \\
A_{32} & = 2 \pi \alpha \psi^5 \left ( 6 + \frac{2}{c_s^2} \right )\,, \nonumber \\
A_{33} & = -\frac{2}{r}\,, \nonumber \\
A_{34} &= 2 \pi \psi^4 (\rho h + 5 P)\,, \nonumber \\ 
A_{36} & = 8 \pi \alpha \psi^4 (\rho h + 5 P)\,, \nonumber 
\end{align}
\begin{align}
A_{43} & = 1\,, \nonumber 
\end{align}
\begin{align}
A_{51} & = 2 \pi \rho h \psi^5  \mathcal{B}\,, \nonumber\\
A_{52} & = -2 \pi \psi^5 \frac{1}{c_s^2}\,,  \nonumber \\
A_{55} & = -\frac{2}{r}\,, \nonumber \\
A_{56} & = -10 \pi \psi^4 e\,, \nonumber \\
A_{65} & = 1\,. \nonumber
\end{align}
%

\section{Alternative numerical method}
\label{sec:appendix:alt}

We have also computed the eigenmodes using an alternative numerical method. 
Instead of solving the system of 6 coupled ODEs, we solve a system of 2 ODEs for the 
fluid variables ($\eta_r$ and $\eta_\perp$ for the case $l\ne0$, or $\eta_r$ and $\delta \hat P$ for $l=0$)
coupled with two elliptic equations for the metric perturbations $\delta \hat \psi$ and $\delta \hat Q$
(Eqs.~(\ref{eq:metric:L1}) and (\ref{eq:metric:L2}) for $l\ne 0$ or Eqs.~(\ref{eq:metric:L01}) and (\ref{eq:metric:L02}) 
for $l=0$). 
 
Each of the metric equations, as well as the corresponding boundary conditions given by Eqs.~(\ref{eq:BCS1})-(\ref{eq:BCS2})
are discretised to second order accuracy and written as two linear systems of equations that are solved using the LAPACK 
library\footnote{Linear Algebra PACKage library, \url{http://www.netlib.org/lapack/}.}. Details on the implementation and tests of the elliptic solver
can be found in \cite{Adsuara:2016}. Since the metric equations and the fluid equations are coupled, we obtain the solution of the system of
four equations (for each value of $\sigma$) in an iterative way. We first integrate the fluid variables considering $\delta \hat Q=\delta \hat \psi=0$ (Cowling), then we use
the values of their values to compute the metric perturbations, and continue with the iteration until the residual of all four quantities, computed
as the L2 norm of the difference between two consecutive iterations is below $10^{-4}$. For most of the values of $\sigma$, this procedure converges to a solution in less than $\sim10$ iterations.
Below a certain threshold in the value of $\sigma$, the iterative procedure becomes unstable and no convergence is achieved, even using a small relaxation
factor in the iteration. The reason is that towards lower values of $\sigma$, there appear g-modes with increasing number of nodes. For sufficiently 
low values of $\sigma$, the g-modes have a number of nodes comparable to the number of grid points and this triggers point-to-point numerical noise, 
which spoils the solution. However, this is not a limitation of our method, because it only affects the calculation of very high-order g-modes 
(typically above order 10), which are not relevant to GW observations \citep[see discussion in Section 5.3 in][]{Torres-Forne:2018}  and 
are not well resolved in simulations, anyway.

We have compared the eigenmodes computed with this alternative method with the ones given in the main text. The discrepancy in the eigenfrequencies 
is in all cases below $0.1\%$. 
Given that the alternative method becomes numerically unstable in some cases and does not improve the solution, 
we only present results obtained with our main method in this work.

\section{Eigenfunction classification procedure}
\label{sec:appendix:class}

We describe in this appendix the steps that we carry out to classify the eigenmodes of our linear analysis. This classification
is based on the similarity with the modes from  previous time steps. 

\begin{enumerate}

\item For each mode at each time, we interpolate linearly the values of $\eta_r$ and $\eta_\perp$ (and all necessary quantities) to an equally spaced grid of 300 points between
the centre and the shock location. Instead of the radial coordinate we use a rescaled coordinate $x\in [0,1]$, which maps the interval $r\in[0,r_{\rm core}]$ to $x \in [0,1/3]$, 
the interval $r \in [r_{\rm core},r_{\rm pns}]$ to $x \in [1/3,2/3]$ and the interval $r \in [r_{\rm pns},r_{\rm shock}]$ to $x \in [2/3,1]$, where $r_{\rm core}$, $r_{\rm pns}$ and
$r_{\rm shock}$ are the radial location of the core surface, the PNS surface and the shock.

\item We normalise the eigenfunctions such that all have the same mean energy density, $E / V = 1$, where $E$ is the energy of the mode, given by Eq.~(\ref{eq:energy2}),
and $V$ is the volume of the region inside the shock. In this way, the eigenfunctions at different times are easier to compare with each other.

\item Next, we count the number of interior nodes (not counting the ones at the centre and at the shock) by searching for changes in the sign of $\eta_r$. This information 
can be used to classify the modes according to the Cowling or ESO classification schemes, and serves as a guide here.

\item We compute the energy density of each mode, $\mathcal{E}(x)$, using Eq.~(\ref{eq:energyden2}). Our matching algorithm is based on the similarity of this function
at different times. This quantity has some advantages with respect to the eigenfunctions itself. Firstly, it is a combination of the radial and perpendicular parts, and secondly
it is a positive function and does not suffer from the sign ambiguity that the eigenfunctions have.

\item Our matching algorithm serves to identify a series of modes at different times as members of the same class. However, it does not give a name for the class. 
We use the ESO classification at the starting point of the algorithm to give a preliminarly tag for the modes.
In most of the cases we applied our matching procedure backward in time, using as starting point the last time output (the exception being the $l=0$ modes).
For each of the identified modes,  $^l m_n$, with $m=\{ f,p,g,h\}$ denoting the possible mode classes,  we create a template $\mathcal E (r; ^l m_l)$ as a basis for comparison.

\item We proceed to the next time output and we compare all the templates with the energy density of the new eigenmodes  $\mathcal E(r;\sigma)$, where $\sigma$ belongs to 
all possible eigenvalues of the new time output.  For all possible values of $\sigma$ and $^l m_n$ (for the same $l$) we compute the L2-norm of the difference
\begin{equation}
L2 (\sigma | ^l m_n) \equiv 
 \sum_{i=1}^{N} \left ( \mathcal E (x_i; ^l m_n) - \mathcal E (x_i; \sigma) \right )^2 \,.
 \end{equation}
Values of $L2 (\sigma|^l m_n)\ll 1$, indicate a good matching between the eigenfunction corresponding to $\sigma$ and the template 
for $^l m_n$.

\item We restrict our comparison to frequencies $\sigma$ which are close to the sequence corresponding to the template $^l m_n$.
To do this, we extrapolate the sequence of the modes already  classified  as $^l m_n$ to the new time and compare this value
with $\sigma$. If the relative difference is larger than a certain threshold ($10-20\%$) we reject this combination as possible. We use linear
extrapolation in this procedure for the model 35OC and constant extrapolation for the s20. The extrapolation function is a least squares
fit to the last $10-20$ points already classified.

\item We order all possible matching combinations of modes in ascending order (better to worse matching) and assign sequentially to
each unclassified mode a class given by the corresponding matching template. Modes already classified are removed 
from the sequence to avoid repetitions. 

\item If there are unclassified modes, we use them to create new templates using the ESO criterium.

\item We update the templates incorporating the information of the newly classified eigenfunction and repeat the process for the next time output.
For p- and f-mode sequences, the template is a mean of the last 10 classified modes. This allows for smooth variations in the form of the eigenfunction 
over time, in particular the location of the nodes. For g-modes, which evolve more slowly in time, we use all previously classified modes.

\end{enumerate}

Once all modes have a preliminary classification, we perform some modifications to improve the matching:

\begin{enumerate}

\item For f- and p-modes, we reorder the frequencies according to the number of radial nodes at each frequency, such that 
higher order modes have more nodes. This solves some misclassification issues of high order p-modes.

\item We retag manually some of the low order modes such that the time evolution of the profiles  of $\mathcal E$ 
is similar to the corresponding modes in the decoupled case (see Section~\ref{sec:match:class}), and that the p-modes are approximately integer
multiples of the f-mode.
 
\end{enumerate}

Note that the classification is not fully automatic as it requires the adjustment of a few parameters and thresholds and the manual retag at the end,
for each of the models.

\bsp	
\label{lastpage}
\end{document}